\documentclass[twocolumn,trackchanges]{aastex701}

\newcommand{\rev}[1]{\color{black}{#1}}

\usepackage{wrapfig}
\usepackage{subfigure}

\begin{document}

\title{An exploration of in-band spectral index techniques for extended radio sources with EMU}

\author[0009-0003-2088-9433]{B. D. Ball}
\affiliation{Department of Physics, University of Alberta, Edmonton, Alberta, T6G 2E1, Canada}
\email[show]{bdball@ualberta.ca}

\author[0000-0001-5953-0100]{R. Kothes}
\affiliation{Dominion Radio Astrophysical Observatory, Herzberg Astronomy \& Astrophysics, National Research Council \\Canada, 717 White Lake Road, Kaleden, BC V0H 1K0, Canada}
\affiliation{Department of Physics, University of Alberta, Edmonton, Alberta, T6G 2E1, Canada}
\email{Roland.Kothes@nrc-cnrc.gc.ca}

\author[0000-0002-5204-2259]{E. Rosolowsky}
\affiliation{Department of Physics, University of Alberta, Edmonton, Alberta, T6G 2E1, Canada}
\email{rosolowsky@ualberta.ca}

\author[0000-0001-7722-8458]{J. L. West}
\affiliation{Dominion Radio Astrophysical Observatory, Herzberg Astronomy \& Astrophysics, National Research Council \\Canada, 717 White Lake Road, Kaleden, BC V0H 1K0, Canada}
\affiliation{School of Natural Sciences, University of Tasmania, PO Box 807, Sandy Bay, TAS 7006 Australia}
\email{}

\author[]{W. Raja}
\affiliation{Australia Telescope National Facility, CSIRO, Space and Astronomy, P.O. Box 76, Epping, NSW 1710, Australia}
\email{}

\author[0000-0002-4990-9288]{M. D. Filipovi\'c}
\affiliation{Western Sydney University, Locked Bag 1797, Penrith South DC, NSW 2751, Australia}
\email{}

\author[0000-0002-6097-2747]{A. M. Hopkins}
\affiliation{School of Mathematical and Physical Sciences, 12 Wally’s Walk, Macquarie University, NSW 2109, Australia}
\email{}

\author[0000-0001-6109-8548]{S. Lazarevi\'c}
\affiliation{Western Sydney University, Locked Bag 1797, Penrith South DC, NSW 2751, Australia}
\affiliation{Australia Telescope National Facility, CSIRO, Space and Astronomy, P.O. Box 76, Epping, NSW 1710, Australia}
\affiliation{Astronomical Observatory, Volgina 7, 11060 Belgrade, Serbia}
\email{}

\author[0000-0002-4814-958X]{D. Leahy}
\affiliation{Department of Physics and Astronomy, University of Calgary, Calgary, Alberta, T2N 1N4, Canada}
\email{}

\author[0000-0001-9414-175X]{S.F. Rahman}
\affiliation{Syed Babar Ali School of Science and Engineering, Lahore University of Management Sciences, Lahore, Pakistan}
\email{}

\author[0009-0009-7061-0553]{Z. J. Smeaton}
\affiliation{Western Sydney University, Locked Bag 1797, Penrith South DC, NSW 2751, Australia}
\email{}

\author[0009-0001-6908-2433]{S. Taziaux}
\affiliation{Ruhr University Bochum, Faculty of Physics and Astronomy, Astronomical Institute (AIRUB), \\Universitätsstraße 150, 44801 Bochum, Germany}
\email{}

\begin{abstract}
Spectral indices are an effective way of differentiating between thermal and nonthermal sources of radio emission. Most often, spectral indices are determined using flux measurements taken over multiple observations with different instruments. Since these data are not always available, determining spectral indices from a single telescope observation is desirable. The reliability of these in-band spectral indices has not been well studied, particularly for extended sources. Here, we explore three different techniques for finding in-band spectral indices with the Evolutionary Map of the Universe (EMU), a radio southern sky survey being conducted with the Australian Square Kilometre Array Pathfinder (ASKAP). These methods include: (1) integrating fluxes in sub-bands, (2) temperature-temperature (T-T) plots, and (3) using the ratio of the Taylor terms produced in the imaging process. We apply these techniques to a selection of extended sources of radio emission, primarily supernova remnants (SNRs) and \ion{H}{2} regions, and assess if the indices produced through these methods agree with the expected values. For the SNRs, we also compare the in-band results to broadband spectral indices determined using flux values obtained from published works. We examine whether and when the in-band techniques can differentiate between thermal and nonthermal sources. In general, we find the T-T plot method to be the most reliable across the range of source sizes and brightnesses studied for both SNRs and \ion{H}{2} regions. The sub-band and Taylor term methods can be useful under certain conditions but cannot be considered as broadly reliable for extended sources.
\end{abstract}

\section{Introduction} \label{sec:intro}

Radio surveys are a powerful tool for studying the Galactic plane, allowing us to peer deep into regions obscured by interstellar dust. Galactic radio emission is primarily produced either through the nonthermal synchrotron process, by supernova remnants (SNRs), or thermally via bremsstrahlung emission, mainly by clouds of ionized hydrogen known as \ion{H}{2} regions. Observable characteristic differences in the emission produced through these mechanisms can be used to determine the origin of the emission and help to classify its source. 

One way to distinguish between thermal and nonthermal emission sources is by measuring their spectral indices. The spectral index ($\alpha$) characterizes a source's spectral energy distribution in the radio and can be defined as \(S(\nu)\propto \nu^{\alpha}\), where $S$($\nu$) is the flux density at a frequency $\nu$. Optically thin thermal bremsstrahlung emission is associated with a flat spectrum, producing an index of $\alpha\simeq-0.1$. As free-free absorption effects become significant, the spectral index becomes increasingly positive, approaching $\alpha=+2$ in the optically thick regime \citep{Longair2011}. Synchrotron radiation produces a steeper negative index that depends on the particle energy distribution and underlying acceleration mechanism \citep{Rybicki1986}. For SNRs, observations show the spectral index is typically found within the range \(\alpha=-0.5\pm0.2\) \citep{Dubner2015}. Optically thick sources of nonthermal radio emission can also have steep positive indices at low frequencies, up to +2.5, when synchrotron self-absorption effects become significant \citep{Longair2011}. For SNRs, these effects are observed at very low frequencies, typically below 100 MHz \citep{Abadi2024}. 

Identifying Galactic radio objects as nonthermal sources is a common approach for confirming SNR candidates as true SNRs. The Galactic SNR population is generally believed to be under-surveyed, and the discrepancy between observed and theoretical Galactic SNR populations \citep{Green2015,Ranasinghe2022,Ball2023} may be at least partly attributed to the inability to confirm the several hundreds of proposed SNR candidates as SNRs \citep{Green2025,Ball2025}. Confirming these sources would represent significant progress toward resolving the Galactic SNR population discrepancy. Part of the challenge in this work lies in the limitations to multi-wavelength studies imposed by the presence of the Galactic plane, meaning that in many cases we must rely solely on radio observations. Sources can be confirmed as SNRs based on their radio emission alone either through the detection of linear polarization or by measuring nonthermal spectral indices. 

Spectral indices are typically determined using flux measurements taken across multiple observations, often with different telescopes, at a wide range of frequencies. Here, we refer to this technique as the broadband method. However, we can also take advantage of the bandwidth of a single observation to determine an in-band spectral index. This approach is most beneficial when multiple high-quality observations of a source are not available, though the limited frequency range of the in-band method introduces larger uncertainties. There are several different approaches that may be used to determine in-band indices, three of which are explored in this paper.

There have been systematic studies of in-band spectral index techniques with point sources \citep{Smolcic2017,Norris2021,Rashid2024,Mangla2025,Thomson2026} and studies of the in-band indices of specific targets of extended radio emission \citep{Bhatnagar2011,Reynoso2015,Ceccotti2023}. However, to our knowledge there has not been a systematic study that compares in-band spectral index techniques for extended sources of radio emission. The latest wave of high-resolution interferometric radio data presents new opportunities for such studies. The current generation of Galactic radio surveys, such as those being conducted with the Australian Square Kilometre Array Pathfinder (ASKAP) and MeerKAT \citep{Hopkins2025,Goedhart2024}, allow us to study the Galactic plane in greater detail than ever before. In this work, we seek to better understand the reliability of in-band spectral indices for extended radio sources using these new surveys. These results may also have implications as we look to the future radio surveys that will be conducted with the Square Kilometre Array (SKA) and next-generation Very Large Array (ngVLA). 

The goal of this paper is to test the performance of three different spectral index techniques for extended sources of radio emission using data from the Evolutionary Map of the Universe (EMU) radio sky survey that is being conducted with ASKAP. The three techniques studied are: (1) splitting the full EMU band into four sub-bands and integrating fluxes in the sub-band images, (2) plotting the intensity values of the pixels in one frequency sub-band versus the values in a second sub-band to produce temperature-temperature (T-T) plots, whose slope can be used to derive an index, and (3) using the ratio of the Taylor terms produced in the multi-frequency synthesis (MFS) imaging process. We apply these techniques to thermal and nonthermal extended sources of radio emission and compare the results to the theoretical and broadband values. 

We describe the radio data used in this study in Section~\ref{sec:obs} and the techniques used to determine spectral indices in Section~\ref{sec:methods}. In Section~\ref{sec:results} we present our results, comparing the spectral indices determined using the three in-band techniques to the theoretical and broadband values. We then discuss these results in Section~\ref{sec:discuss}, assessing the performances of the three techniques and examining the conditions under which they are reliable. Finally, in Section~\ref{sec:conc} we summarize our findings.

\section{Observations} \label{sec:obs}

\subsection{EMU Observations with ASKAP}

EMU is an ongoing radio continuum survey being conducted with ASKAP that will eventually cover the entire southern sky \citep{Hopkins2025,Norris2021}. The EMU survey is centred at 943 MHz with a bandwidth of 288 MHz. Each Stokes I field is produced as a mosaic of 36 beams, convolved to a resolution of 15$''$ with a median rms noise of 30 $\mu$Jy beam$^{-1}$ \citep{Hopkins2025}. EMU data is processed using the standard \textsc{ASKAPsoft} pipeline \citep{Guzman2019}, including MFS imaging, multiscale cleaning, and self-calibration \citep{Norris2021,Hopkins2025}. We use data from many different EMU fields that cover the Galactic plane, most of which are listed in \cite{Ball2025}. The data used are available on the CSIRO ASKAP Data Science Archive (CASDA)\footnote{\url{https://research.csiro.au/casda/}}.

In addition to the processed Stokes I images, EMU provides a selection of associated datasets that are also available on CASDA. This includes the visibilities, which we use in our sub-band integration technique, reprocessing these data to produce our sub-band images. The details of this procedure are further described in Section~\ref{sec:sub-band}. The MFS imaging process, included in the ASKAP pipeline, uses a Taylor term expansion to account for spectral changes in the source \citep{Rau2011}. The first- and second-order Taylor term maps, TT0 and TT1, respectively, are produced by EMU for each tile and are also available on CASDA. We use these maps to determine our Taylor term indices.

The EMU survey is commensally observed with the Polarization Sky Survey of the Universe’s Magnetism (POSSUM) \citep{Gaensler2025}, a survey of the polarization properties of the same region. The POSSUM data products include continuum cubes for all Stokes parameters with 1 MHz channel resolution. The 288 channel Stokes I cubes can be alternatively used to obtain sub-band indices, rather than reprocessing the visibilities to form sub-band images. We discuss this possibility further in Section~\ref{sec:sub-band}.

\subsection{Ancillary Radio Data}

We use data from other radio surveys to determine flux densities that can then be used to calculate broadband spectral indices for the SNRs. The Southern MeerKAT Galactic Plane Survey (SMGPS) is a radio continuum survey of the Galactic plane conducted by MeerKAT at a central frequency of 1360 MHz \citep{Goedhart2024}. We use data from the SMGPS to determine flux densities for sources that are covered in the survey. High-latitude sources were not included as the SMGPS only covers the Galactic plane.

We also use data from the GaLactic and Extragalactic All-sky MWA (GLEAM) survey \citep{HurleyWalker2019_survey} to calculate low-frequency flux densities. There are four different GLEAM bands centred at 88 MHz, 118 MHz, 155 MHz, and 200 MHz. The GLEAM surveys have relatively poor resolution and sensitivity when compared to EMU and the SMGPS so flux densities could only be determined if the source was large and bright enough to be visible. Several sources were visible in some of the GLEAM bands but not others. The fluxes determined using the SMGPS and GLEAM data were used in combination with the EMU fluxes and published results to determine broadband spectral indices that can be compared to the indices determined with the in-band techniques. 

\section{Methods} \label{sec:methods}

The goal of this paper is to evaluate the reliability of three different in-band spectral index techniques for extended sources with EMU data. Here we outline the in-band techniques used to determine spectral indices. We also describe the method for determining the broadband indices, which were used as a comparison point for the in-band values, and the flux integration process. 

\subsection{In-band Index Techniques}

\subsubsection{Sub-band Integration}\label{sec:sub-band}

The sub-band spectral indices are determined by splitting the 288~MHz wide frequency band that EMU observes over into four 72~MHz wide sub-bands using the ASKAPsoft program \textsc{mssplit}. The decision to use four sub-bands was based on the inherent trade-off between establishing a well-defined linear relation, which benefits from more data points, and maximizing the signal-to-noise ratio (S/N), which improves with wider sub-bands. Imaging is then performed on each of the sub-bands using the standard ASKAPsoft pipeline \citep{Guzman2019}. Each EMU tile is a mosaic of 36 beams and the sub-band imaging process is performed individually on a beam-by-beam basis. Thus, we do not process entire fields and only perform this sub-band imaging for beams that cover sources of interest. The new sub-band images have a resolution of 18$''$. 

Integrated flux densities and corresponding uncertainties are determined for the sources in each of the four sub-bands, centred at 836 MHz, 908 MHz, 980 MHz, and 1052 MHz, using the technique described in Section~\ref{sec:flux}. The spectral indices can then be calculated using a linear fit in log space through the four flux values. To estimate index errors, we apply a bootstrapping technique, repeatedly resampling the four flux values from within the range defined by their measurement uncertainties. The spectral indices and errors listed are determined as the mean and standard deviation calculated from 20,000 of these samplings. It should be noted that the derived uncertainties may be underestimated, as this analysis only accounts for measurement errors and does not capture potential systematic errors.

We also use the sub-band images to produce spectral index maps by performing a pixel-by-pixel linear fit between $\ln S(\nu$) and $\ln \nu$ for the four sub-bands. The spectral index maps presented here have been convolved to a slightly lower resolution than the original images to smooth out some of the noise while aiming to preserve any real structure in the index maps. These maps can provide insight into changes in the spectral index across the source. In some sources, the index maps reveal that the indices may be accurate for small-scale features but not for the source as a whole.

It is also possible to apply the sub-band integration technique to the POSSUM Stokes I data cubes, rather than re-imaging the data into four sub-band images. This can be done either by fitting a spectrum to the 288 1 MHz channels or by averaging these 1 MHz channels into wider sub-bands, instead of producing them through re-imaging. In this work, we chose to re-image the data in an attempt to produce cleaner sub-band images with lower noise, which should allow us to apply this technique to fainter sources. We do not explore these other sub-band options in great depth, but do include a brief comparison using the studied SNRs to analyze the potential value provided by the re-imaging process. These findings are discussed in Section~\ref{sec:subband_options}.

\subsubsection{T-T Plots}

In-band spectral indices can also be determined using T-T plots \citep{Turtle1962}. \rev{The T-T plot method relies on combining information from many pixels to obtain a single spectral index value, thereby increasing the S/N at the expense of spatial information.} For this method, we use the four sub-band images that were produced for the sub-band integration technique. For each combination of the four sub-bands, we graph the intensity values ($S_1$) of every pixel in a region of interest at one sub-band frequency ($\nu_1$) versus the intensity values ($S_2$) of those pixels at a second sub-band frequency ($\nu_2$). The source boundaries were manually defined and are the same as those used for the sub-band flux density calculations. For a source where the index is relatively constant, this should produce a linear relationship with the slope being equivalent to \(S_2/S_1\). This ratio can then be used to derive the spectral index: \[\alpha=\frac{\ln(S_2/S_1)}{\ln(\nu_2/\nu_1)}.\] 

12 different T-T plots can be produced if we use all combinations of the four sub-bands on each axis. In practice, we use a linear fit that is weighted based on the intensity value to place more importance on the brighter pixels, which are typically fewer in number but dominate the emission. The index values we provide here are the mean results from all 12 combinations and the index uncertainties are the standard deviations. We find that uncertainties for individual T-T plots are generally insignificant (on the order of 0.001) compared to the uncertainties derived from the differences between T-T plots.

\subsubsection{MFS Taylor Terms}

In-band spectral indices can also be derived using the Taylor terms that are produced in the MFS imaging process \citep{Rau2011,Norris2021}. In this process, the intensity for each pixel is modeled by a Taylor series expansion about a reference frequency, $\nu_0$. From this we can write
\[S(\nu)=S(\nu_0)(\frac{\nu}{\nu_0})^{\alpha+\beta \log(\nu/\nu_0)}\]
where $\alpha$ is the spectral index and $\beta$ represents the spectral curvature. The EMU survey produces TT0 and TT1 coefficient maps based on the first two terms of this Taylor series. 

While the introduction of higher-order terms encodes information about the spectral curvature, it can also introduce larger errors, particularly for low-S/N sources \citep{Urvashi2010}. For EMU, the possibility of including the second-order Taylor term was explored in the pilot survey, but it was found that few sources had observable curvature within the 288 MHz bandwidth and that the introduction of this term resulted in increased uncertainty without increasing accuracy \citep{Norris2021}. Here we adopt the reasonable assumption that our spectra are linear within the EMU bandwidth ($\beta$=0).

We use the TT0 and TT1 maps from EMU to produce a spectral index map by taking a pixel-by-pixel ratio of the Taylor terms with $\alpha$ = TT1/TT0. We also find an average index for the whole source by summing the pixels from the TT1 map and dividing by the sum of the pixels from the TT0 map. To calculate uncertainties, we measure the rms noise in the region surrounding the source and propagate this value forward, applying a bootstrapping technique similar to that used for the sub-band uncertainties. The mean and standard deviation from 10,000 samples are taken as the Taylor term indices and corresponding errors. This technique relies solely on data that are available on CASDA and thus is the quickest and simplest to implement of the in-band techniques.

\subsection{Broadband Indices}

We determine broadband spectral indices for the SNRs studied utilizing the flux values we calculate from EMU, the SMGPS, and GLEAM in combination with data obtained from published works. The broadband indices are determined as a comparison point for the in-band values to assess the reliability of the different techniques, though they should not be considered as definitive values. The previously published flux values are taken from the \cite{Green_cat} catalogue and references therein. A full list of references for the fluxes used is included in Table~\ref{tab:known_snrs}. For the broadband index calculations, we aim to include as many flux data points as possible but reject any that deviate significantly from the spectrum suggested by the other values. This most commonly occurs in large sources that are clearly missing flux in interferometer data. 

\subsection{Flux Integration} \label{sec:flux}

Flux integration is performed using a modified version of the \textsc{Polygon\_Flux} software developed by \cite{HurleyWalker2019} using the technique described in \cite{Ball2023}. The software integrates the flux within a defined region, performs background subtraction, and allows the user to remove bright point sources. We use this software to determine the 943 MHz flux densities listed in Table~\ref{tab:known_snrs}, some of which have been previously listed in \cite{Ball2025}, as well as the SMGPS and GLEAM values that were used in the broadband spectral index determination. Uncertainties are based on changes in the flux value corresponding to slight variations in background sampling and source perimeter definition. We also use the region defined for the flux integration to determine the source sizes that are listed in Table~\ref{tab:known_snrs}. The sizes provided are based on the diameter of a circle that would have the same area as the region of integration.

For the sub-band integration technique, we also use a modified version of the \textsc{Polygon\_Flux} software \citep{HurleyWalker2019} to determine flux densities in each of the four sub-bands. This modified version ensures the integration is performed over the exact same region for each of the bands, with the same background sampling and point source removal. 

\section{Results} \label{sec:results}

We assess the reliability of the three in-band spectral index techniques by comparing the values determined with these techniques to the theoretical and broadband values for different types of extended sources. Most of the sources studied are known Galactic \ion{H}{2} regions or SNRs. \ion{H}{2} regions produce radio emission thermally and are expected to have flat ($-$0.1) or positive (up to +2) spectral indices. SNRs produce nonthermal radio emission and generally should have indices within the range \(-0.7\leq\alpha\leq-0.3\). We also show results from one extragalactic source, the radio galaxy Centaurus B, as a test source for the in-band techniques. 

\subsection{Centaurus B (B1343$-$601)}

\begin{figure*}
    \subfigure{\includegraphics[width=0.5\linewidth]{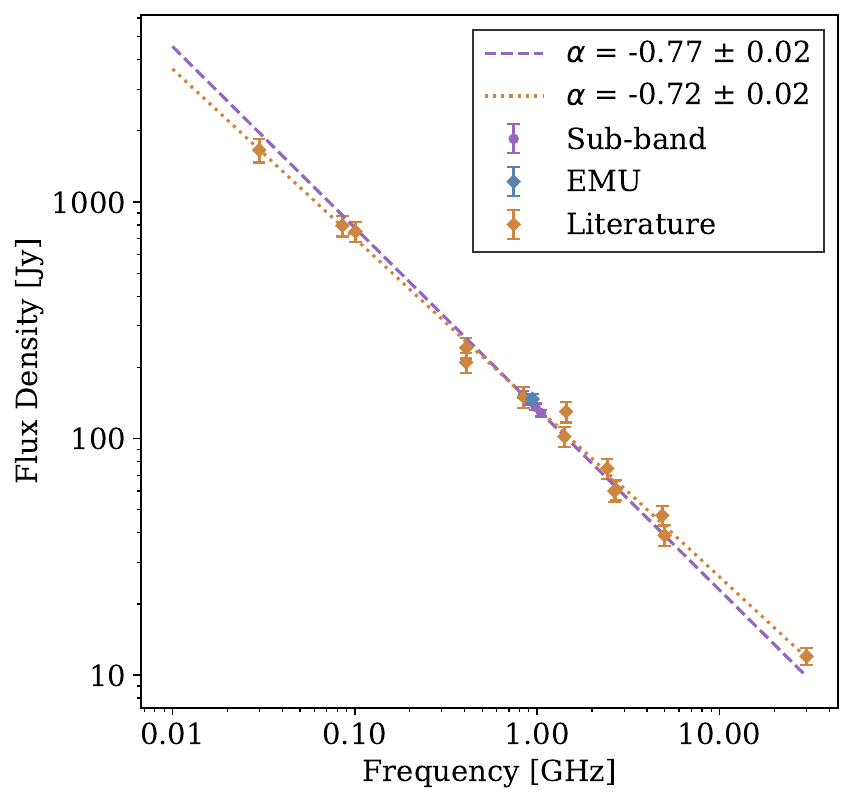}}\hfill
    \subfigure{\includegraphics[width=0.48\linewidth]{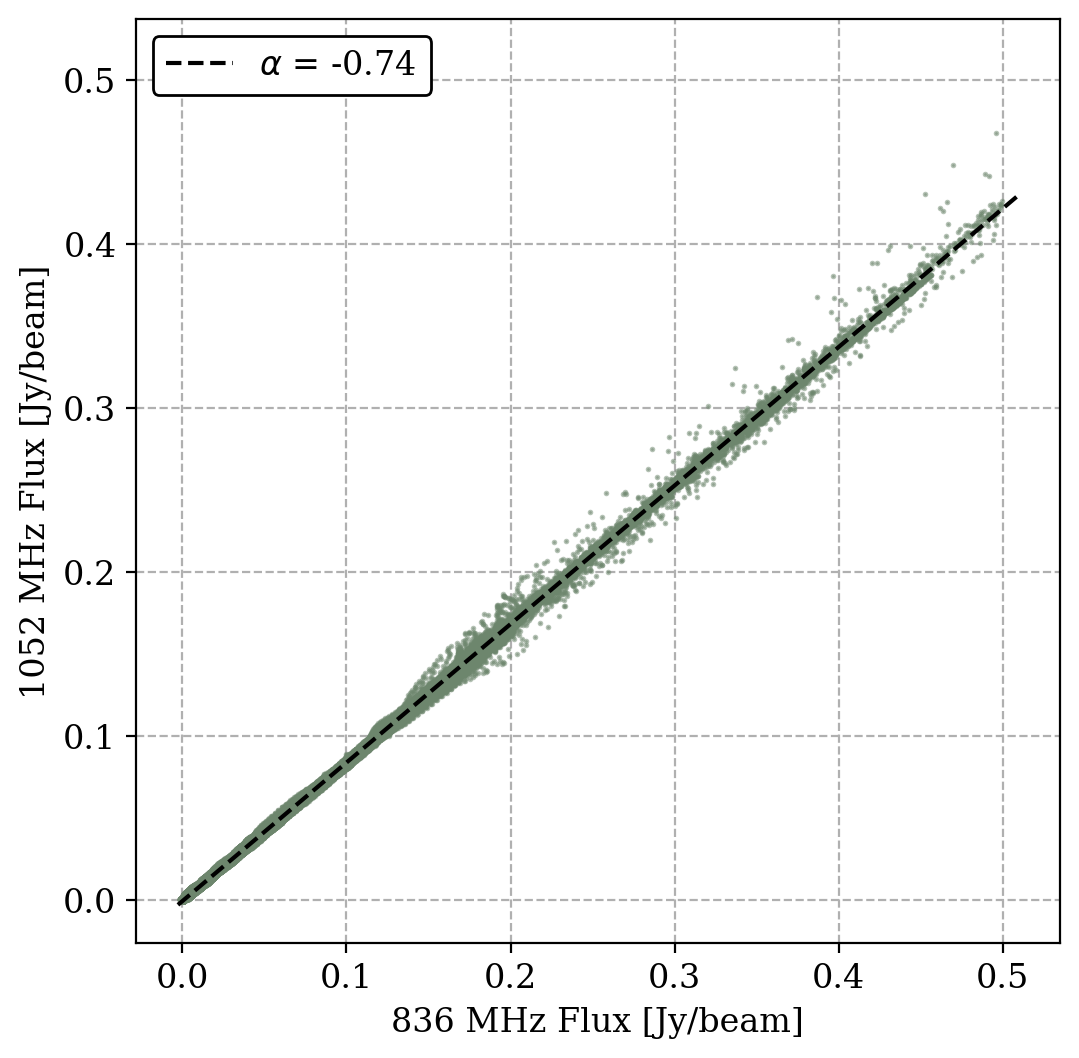}}
    \caption{The left figure compares the broadband and sub-band spectral indices for Cen B. The broadband value is determined using published flux data in combination with the flux value from EMU. The sub-band index is determined using the flux densities in the four sub-bands. The right figure shows a sample T-T plot for Cen B. Flux values from the lowest-frequency sub-band for each pixel are represented on the x-axis and flux values from the highest-frequency sub-band are shown on the y-axis. The slope of the linear fit is then used to determine $\alpha$.}
    \label{fig:cenB2}
\end{figure*}

\begin{figure*}
    \centering
    \subfigure{\includegraphics[width=0.9\linewidth]{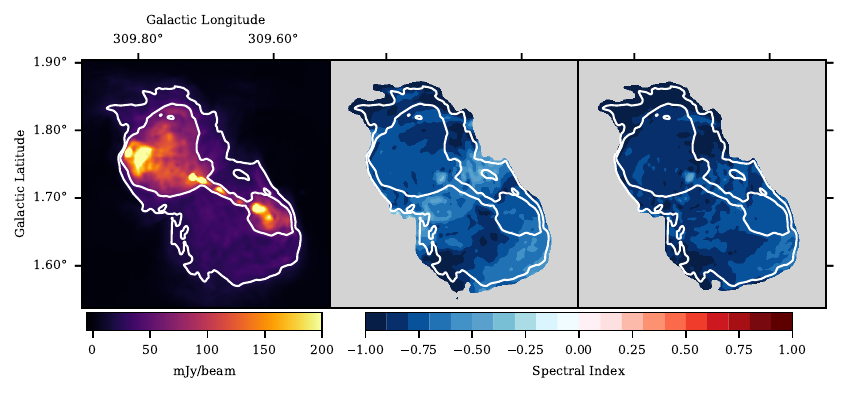}}
    \caption{Images of Cen B. Here we show Stokes I from EMU at 943 MHz (left), the sub-band spectral index map (middle), and the Taylor term spectral index map (right).}
    \label{fig:cenB}
\end{figure*}

The extragalactic radio sky is populated by radio galaxies, sources with extended emission powered by jets from active galactic nuclei. This radio emission is produced through the synchrotron process, though the spectra of radio galaxies are generally slightly steeper than what is expected for SNRs. Radio galaxies have typical spectral indices of around \(-0.8\) and are usually found within the range \(-1.3 \leq \alpha \leq -0.5\) \citep{Kellermann1988}.

The radio galaxy Centaurus B (Cen B) is one of the brightest radio objects in the southern sky, first catalogued by \cite{Mills1952}. It is nearby \citep[$z=0.01215$;][]{Jones2001} and close to the Galactic plane at $b=1.7^\circ$. As such, it is well documented and has been observed across a wide range of frequencies, making it a good test source for our spectral index methods. 

Here, we compare the results from our three in-band spectral index techniques with the broadband value determined by combining the 943 MHz value from EMU (\(147\pm3\) Jy) with published flux values from \cite{Persic2019}, \cite{Jones2001}, and references therein. From this, we obtain a broadband index of \(-0.72 \pm 0.02\), consistent with the value reported in previous work \citep[$-0.73$;][]{Jones2001}. With the T-T plot method, we find an index of \(-0.73\pm0.04\), which agrees well with the broadband value. The sub-band integration index, \(-0.77\pm0.02\), is slightly steeper but also in fairly good agreement, especially considering the limited frequency range. These results are shown in Figure~\ref{fig:cenB2}. The Taylor term approach produces the steepest index at \(-0.83\pm0.04\). This may be, at least in part, because this method does not include background subtraction.

We also produce spectral index maps using the four sub-band channels and the Taylor term coefficient maps. These are shown in Figure~\ref{fig:cenB}. There are noticeable ripples in the index map produced by instrumental effects from the bright point source near the center of the galaxy. However, the integrated flux densities appear to be unaffected and produce an index that is consistent with previous work. The bright central point source appears to have a flatter index than the rest of the galaxy. The spectral index maps also reflect that the Taylor term method produces a steeper index than the sub-band method.

The results from the in-band techniques are remarkably consistent with the broadband value, despite the relatively large size (21$'$) of the source. This is likely because most of the emission comes from small-scale structures, and thus the missing short spacings do not significantly affect the resultant flux densities. The results from Cen B demonstrate that these in-band techniques can produce indices consistent with broadband studies for an extended source in the ideal case where the source is extremely bright, there is not significant missing flux, and where the S/N is very low. 

\subsection{\ion{H}{2} Regions}

\begin{figure*}
    \centering
    \includegraphics[width=0.8\linewidth]{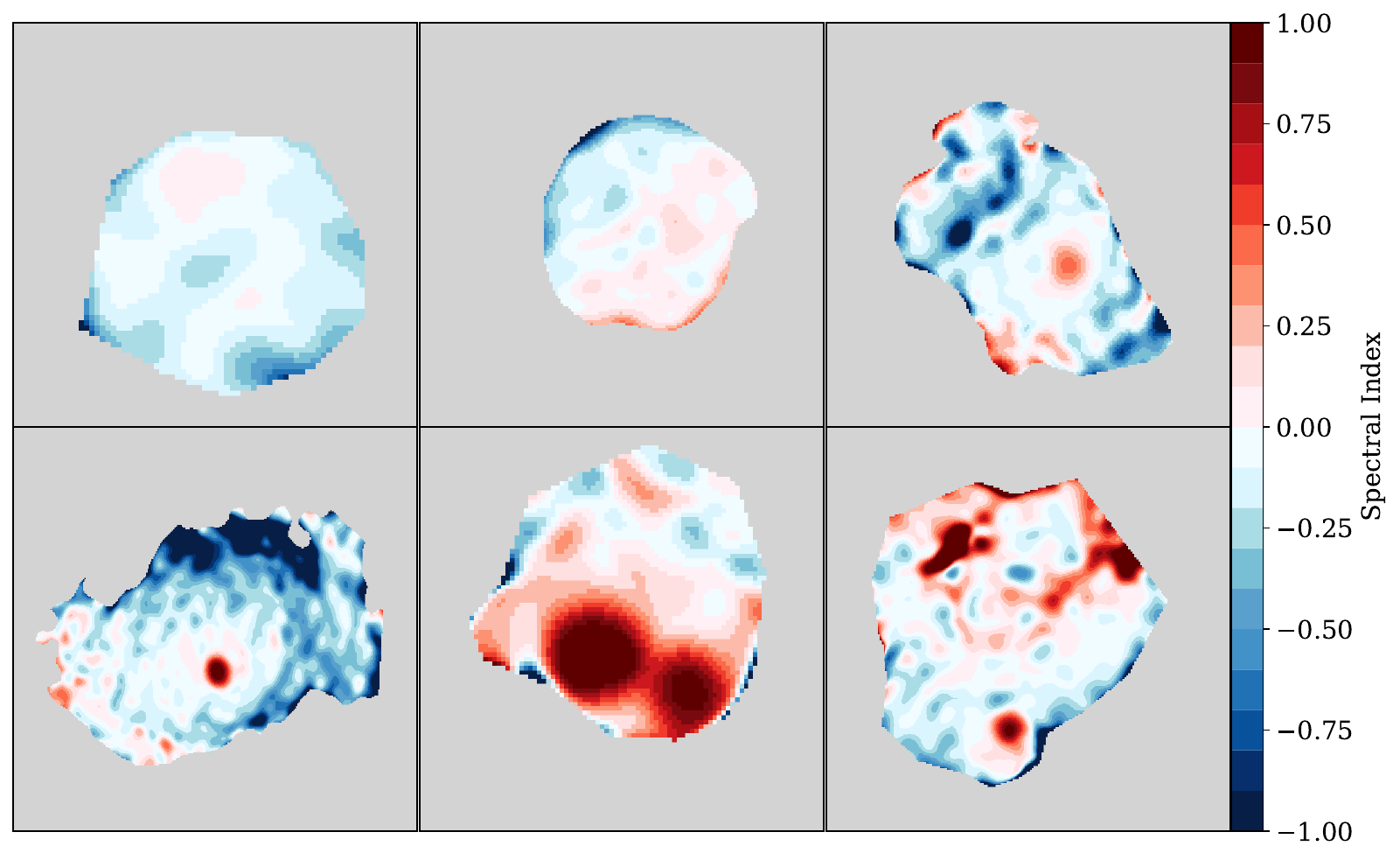}
    \caption{Sample sub-band spectral index maps for six of the \ion{H}{2} regions studied. We can see a mix of optically thin (flat) components and optically thick (steep positive) components. We often find small bright optically thick cores surrounded by optically thin emission. These sources will have overall indices that are an average of these components. Steep negative edges can be seen in some of the sources, corresponding with regions where the emission becomes fainter and the indices are less reliable.}
    \label{fig:hiis}
\end{figure*}

\begin{figure*}
    \subfigure{\includegraphics[width=0.51\linewidth]{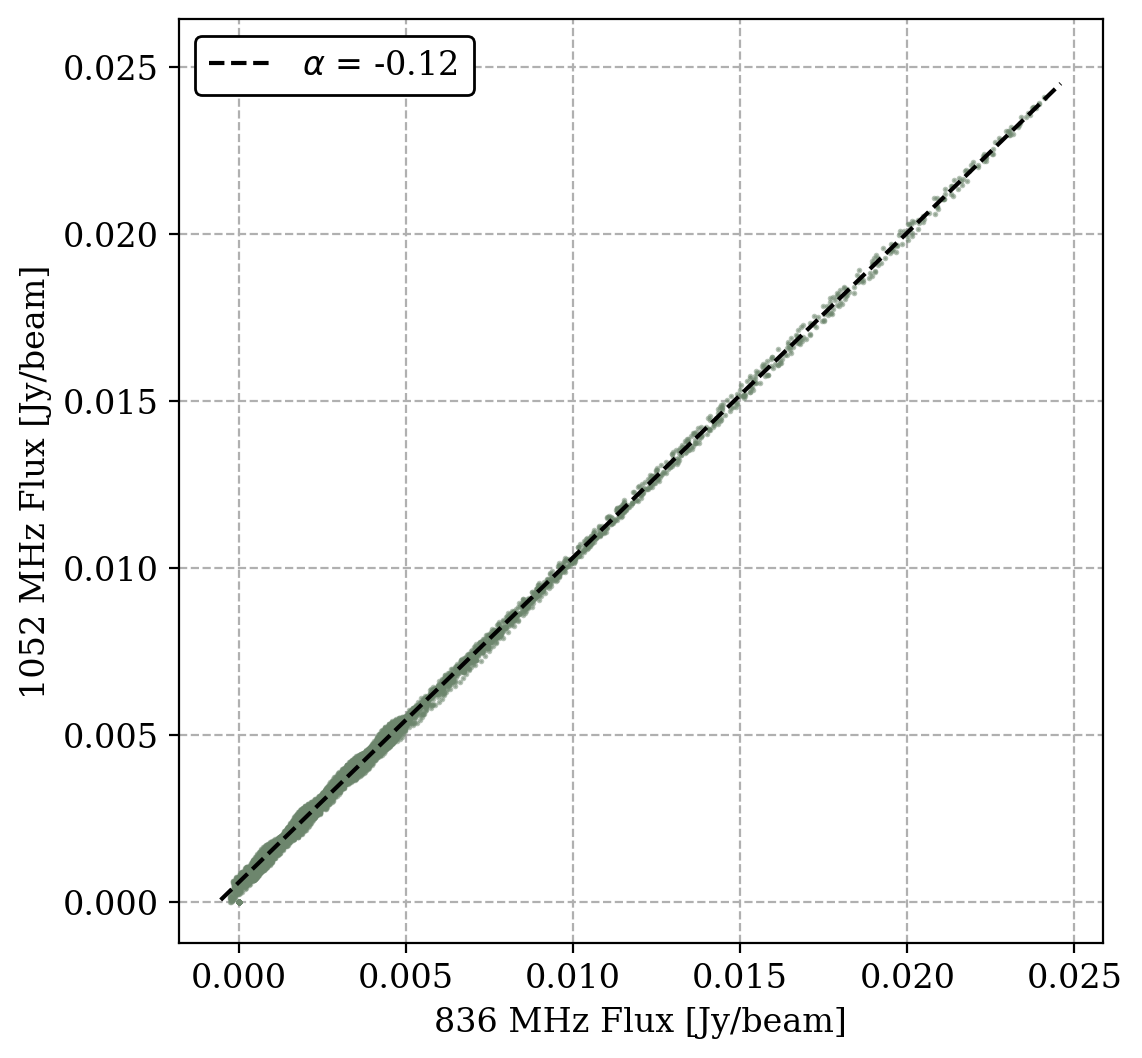}}\hfill
    \subfigure{\includegraphics[width=0.48\linewidth]{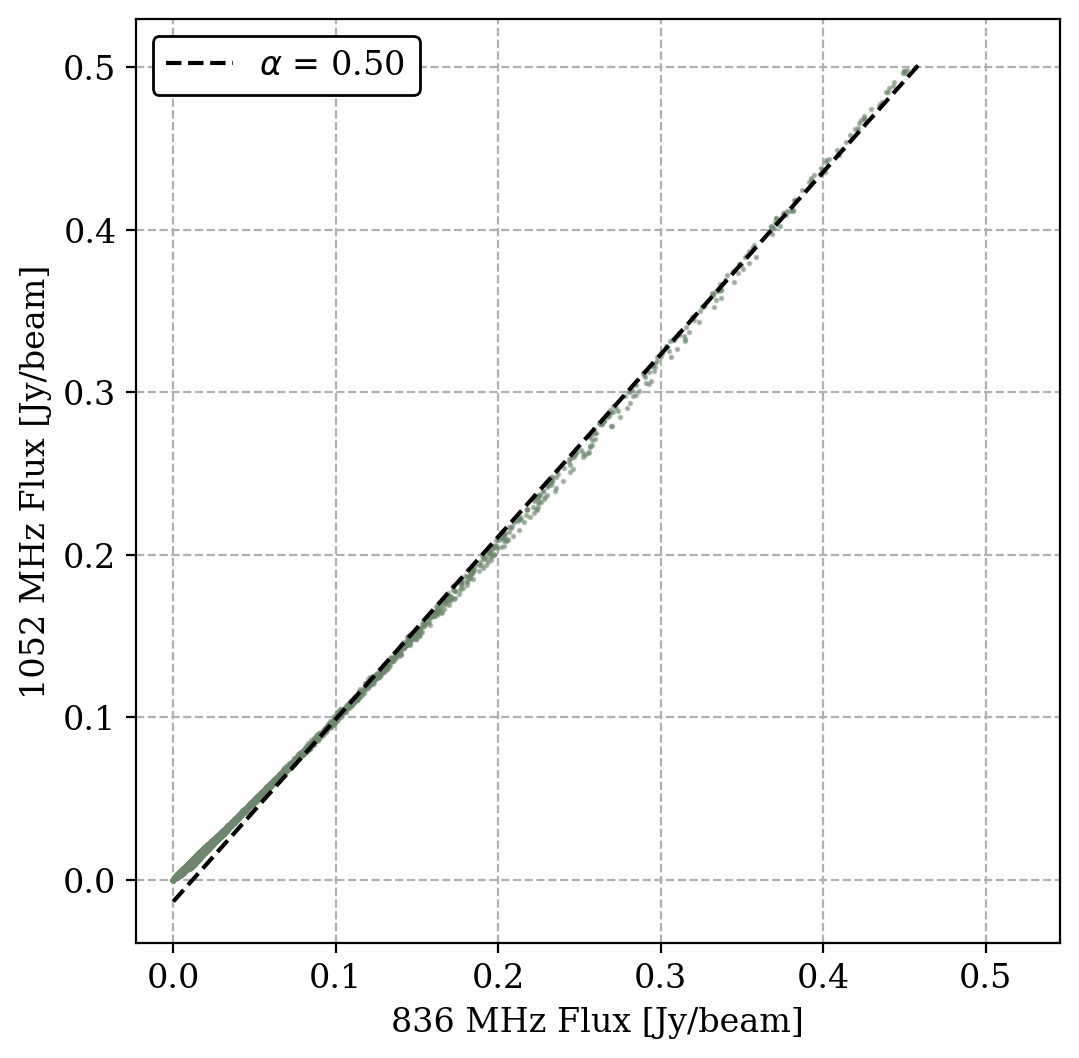}}
    \caption{Sample T-T plots for two \ion{H}{2} regions created using the lowest- and highest-frequency sub-bands. The left figure shows a T-T plot for an optically thin \ion{H}{2} region where we see no significant change in index across the source. The right figure shows a T-T plot for an \ion{H}{2} region with optically thin and optically thick components. There is a small curve in the spectrum and deviation from the linear fit corresponding to the different components: a brighter optically thick core surrounded by fainter optically thin emission.}
    \label{fig:hii_tt}
\end{figure*}

\ion{H}{2} regions should have a spectral index of $-$0.1 if the source is optically thin or a positive index up to +2 if the source is optically thick. We often see both optically thin and optically thick components in a single source, resulting in intermediate indices that are an average of the different components. Here we study a sample of 75 \ion{H}{2} regions taken from the WISE \ion{H}{2} region catalogue \citep{Anderson2014}. We selected sources from the catalogue that are well observed in the EMU data and that span a range of sizes (diameters of 1$'$--25$'$) and brightnesses (fluxes of 0.02--225 Jy). When possible, we chose sources where the boundaries are well defined so it is easier to integrate fluxes over the entire source region. 

Sample sub-band spectral index maps for six \ion{H}{2} regions are shown in Figure~\ref{fig:hiis}. Some of these maps show optically thin sources with flat spectral indices across the whole source. Others show steep positive regions, compact optically thick cores, surrounded by flatter optically thin regions. The same effect can be seen in the T-T plots where we observe a break or curve in the spectrum as a result of the brighter emission coming from the optically thick, steeper component  and the fainter emission coming from the optically thin, flatter component. An example of this is shown in Figure~\ref{fig:hii_tt}.

\begin{figure*}
    \centering
    \subfigure{\includegraphics[width=0.49\linewidth]{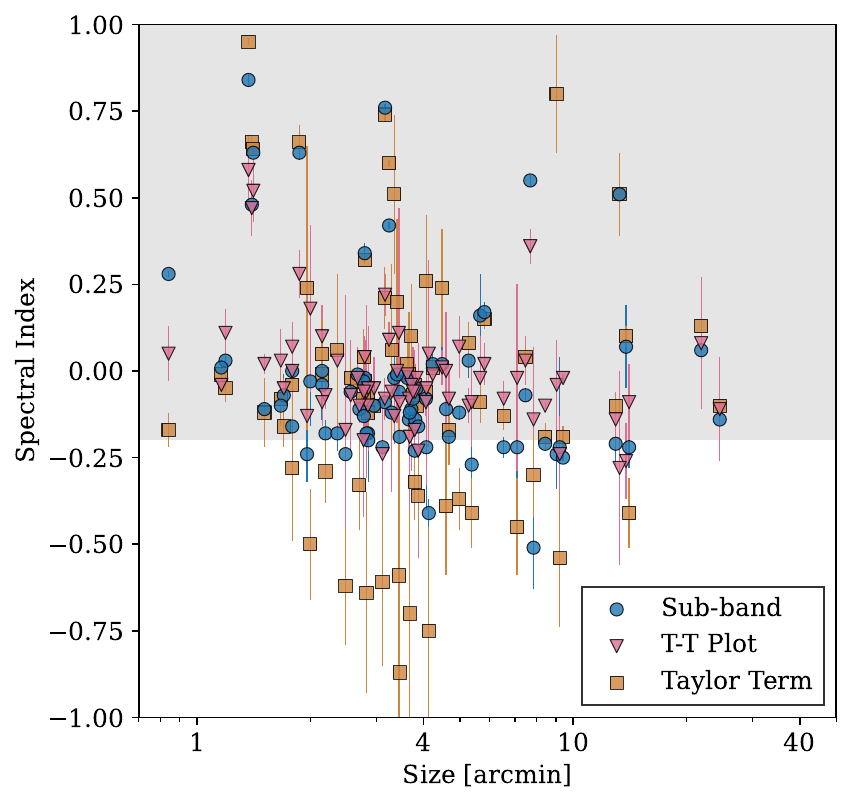}}
    \subfigure{\includegraphics[width=0.49\linewidth]{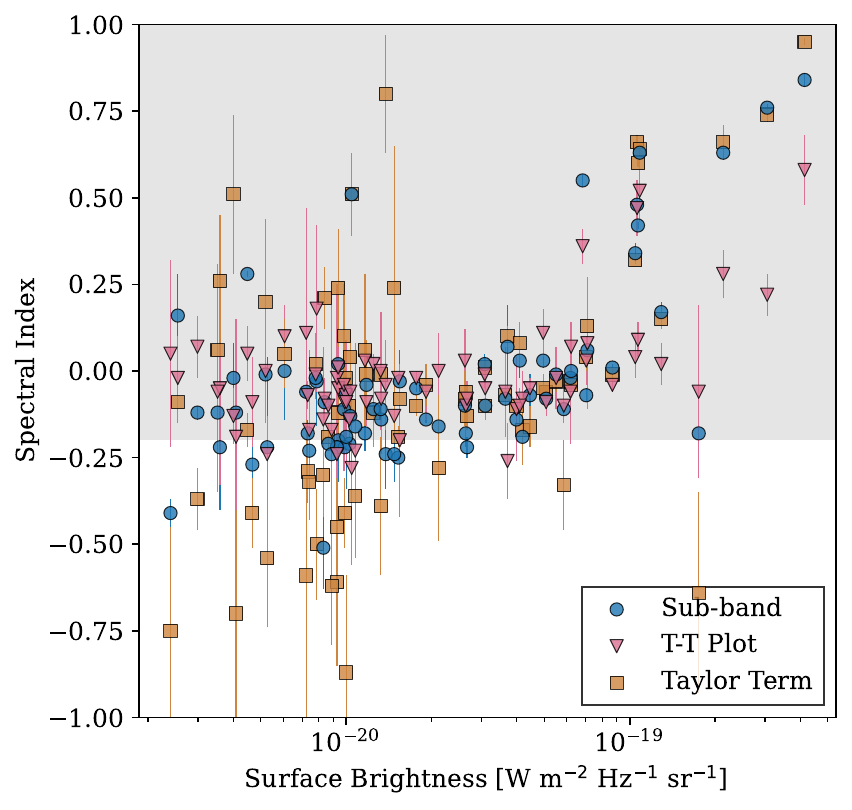}}
    \caption{Spectral indices of a sample of known \ion{H}{2} regions determined using three different in-band techniques shown as functions of angular size and surface brightness. The shaded region reflects the range of expected index values for \ion{H}{2} regions. While almost all of the sub-band and T-T plot indices fall within the expected value range, a significant portion of the Taylor term indices fall below this range. This deviation seems to become more significant with decreasing surface brightness.}
    \label{fig:hii_indices}
\end{figure*}

Because there is a large, diverse sample of known \ion{H}{2} regions, they can be used as a test of reliability for our in-band spectral index techniques. Thus, rather than conducting detailed studies of individual sources, we examine a large sample of \ion{H}{2} regions and present the resultant indices as functions of source size and surface brightness. The results from the three in-band techniques are presented in Figure~\ref{fig:hii_indices}, where the shaded regions indicate the range of expected values for spectral indices of \ion{H}{2} regions. 

From Figure~\ref{fig:hii_indices}, we see that the sub-band and T-T plot methods produce results that are mostly consistent with the theoretical value range, while the Taylor term indices noticeably diverge. Though many of the Taylor term indices are within the expected value range, a significant portion fall below this limit. This deviation seems to be correlated with a decrease in surface brightness but is not obviously correlated with source size. With the sub-band technique, a smaller subset of the values fall outside of the expected value range. This divergence may be correlated with increasing size and decreasing surface brightness, though the effect is not significant within the parameter space studied. The T-T plot values are largely consistent with theoretical expectations, though we do note an increase in uncertainties with decreasing surface brightness. 

In some cases, disagreement between the in-band values may be related to differences in how the techniques weight emission from the optically thin and optically thick components of the sources. This likely explains the differences in index values we see in some of the brightest \ion{H}{2} regions. Here, we see the Taylor term and sub-band indices are generally in agreement and appear to be weighted more strongly toward the brightest parts of the source where the index is steeply positive. The T-T plot indices place more weight on the fainter parts of the source where the index is flatter, resulting in a flatter overall index.

\subsection{Supernova Remnants}

\begin{figure*}
    \centering
    \subfigure{\includegraphics[width=0.49\linewidth]{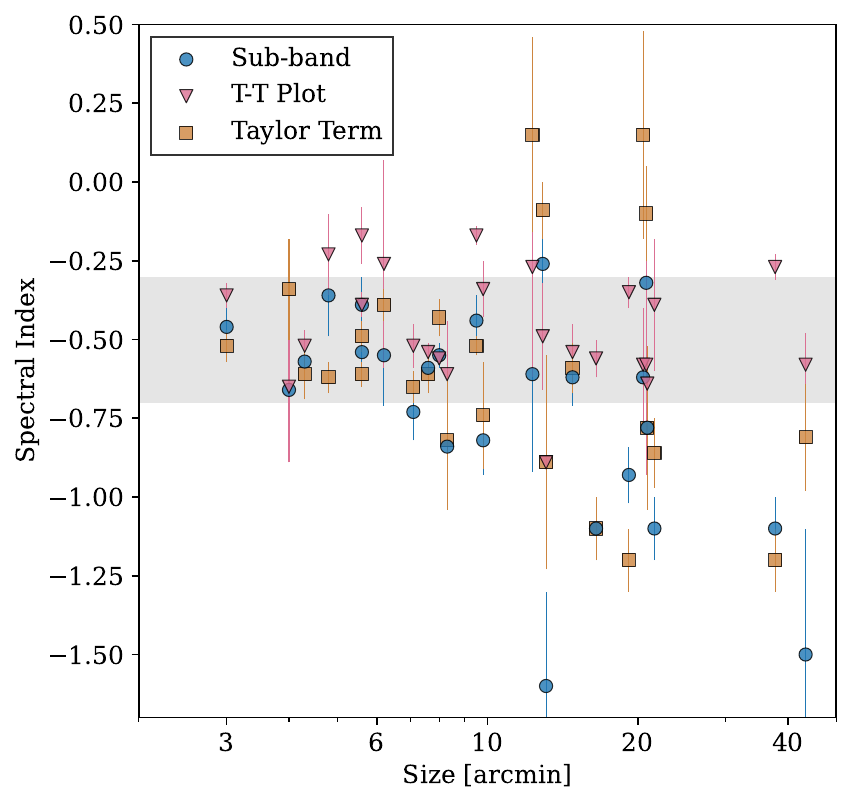}}
    \subfigure{\includegraphics[width=0.49\linewidth]{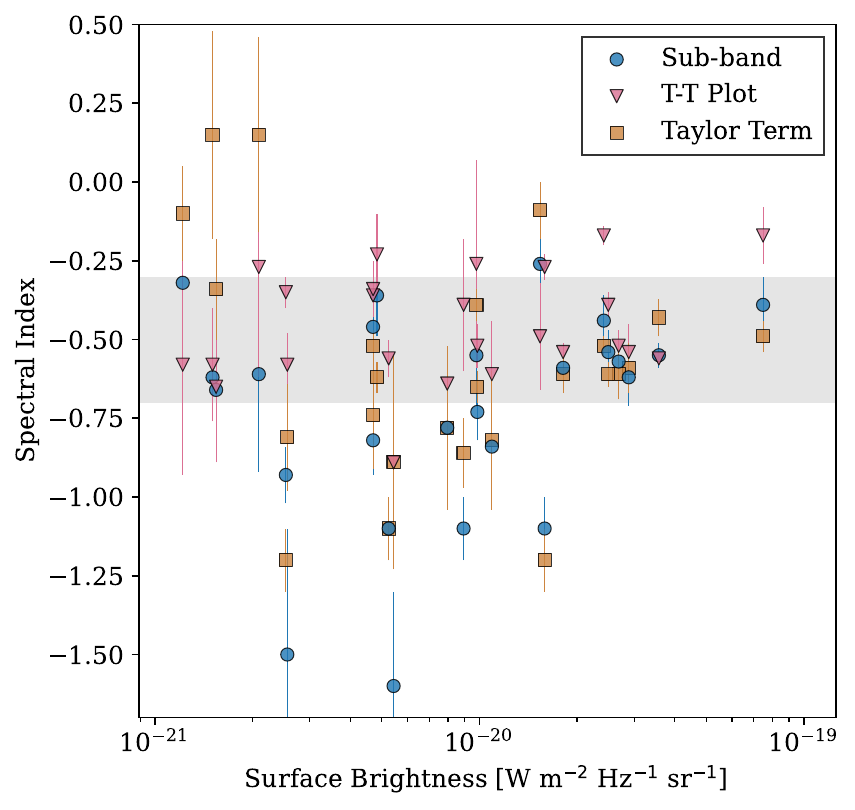}}
    \caption{Comparing the in-band spectral index methods for the studied sample of SNRs as functions of angular size and surface brightness. The shaded region indicates the expected index range for shell-type SNRs. For the sub-band and Taylor term methods, we observe a divergence from the expected value range with increasing source size and decreasing surface brightness. We do not observe this same trend for the T-T plot indices, though we do see an increase in uncertainties with decreasing surface brightness.}
    \label{fig:snr_indices}
\end{figure*}

We study the spectral indices of a selection of 25 known SNRs. The sample is chosen based on which sources have been covered by the EMU survey and for which we can properly integrate fluxes. Very faint sources and sources that are obviously affected by significant instrumental effects were not included. 

Data for the SNRs studied can be found in Table~\ref{tab:known_snrs}. The sizes and flux densities provided are based on EMU data and were determined using the method described in Section~\ref{sec:flux}. The catalogue indices are those listed in the Green catalogue \citep{Green_cat}. The techniques used to determine the broadband, sub-band, T-T plot, and Taylor term spectral indices are described in Section~\ref{sec:methods}. References for the published flux values that were used to determine the broadband indices are included in the table caption. \rev{Detailed descriptions and index maps for a subset of the SNRs studied are included in Appendix~\ref{app:a}.}

\begin{table*}[]
\centering
    \begin{tabular}{c|c|c|c|c|c|c|c|c}
        \hline
         SNR & Size & Flux Density & Flux & Catalogue & Broadband & Sub-band & T-T Plot & Taylor Term\\
         & $[']$ & (943 MHz) [Jy] & Ref. & $\alpha$ & $\alpha$ & $\alpha$ & $\alpha$ & $\alpha$\\
        \hline
        G290.1$-$0.8 & 14.8 & 42 $\pm$ 1 &  1-7 & $-$0.4 & \(-0.45 \pm 0.03\) & \(-0.62 \pm 0.09\) & \(-0.54 \pm 0.09\) & $-$0.59 $\pm$ 0.08\\
        G292.0+1.8 & 9.5 & 14.5 $\pm$ 0.2 & 2,8-10 & $-$0.4 & \(-0.35 \pm 0.03\) & \(-0.44 \pm 0.08\) & \(-0.17 \pm 0.03\) & $-$0.52 $\pm$ 0.03\\
        G293.8+0.6 & 20.5 & 4.2 $\pm$ 0.1 & 6,11 & $-$0.6?  & \(-0.47 \pm 0.16\) & \(-0.62 \pm 0.09\) & \(-0.58 \pm 0.18\) & +0.15 $\pm$ 0.33\\
        G296.8$-$0.3 & 16.5 & 9.5 $\pm$ 0.2 & 6,12,13 & $-$0.6 & \(-0.63\pm 0.04\) & \(-1.1 \pm 0.1\) & \(-0.56 \pm 0.06\) & $-$1.1 $\pm$ 0.1\\
        G304.6+0.1 & 8.0 & 15.2 $\pm$ 0.3 & 2-4,6,14 & $-$0.5 & \(-0.44\pm0.03\) & \(-0.55 \pm 0.04\) & \(-0.56 \pm 0.02\) & $-$0.43 $\pm$ 0.06\\
        G306.3$-$0.9 & 4.0 & 0.164 $\pm$ 0.003 & 15 & $-$0.5? & \(-0.31 \pm 0.04\) & \(-0.66 \pm 0.14\) & \(-0.65 \pm 0.24\) & $-$0.34 $\pm$ 0.16\\
        G309.2$-$0.6 & 13.1 & 6.2 $\pm$ 0.1 & 6,11,13 & $-$0.4? & \(-0.42 \pm 0.04\) & \(-1.6 \pm 0.3\) & \(-0.89 \pm 0.13\) & $-$0.89 $\pm$ 0.34\\
        G310.6$-$0.3 & 8.3 & 5.0 $\pm$ 0.1 & 6 & N/A & \(-0.50\pm0.04\) & \(-0.84 \pm 0.20\) & \(-0.61 \pm 0.17\)& $-$0.82 $\pm$ 0.22\\
        G310.6$-$1.6 & 3.0 & 0.281 $\pm$ 0.003 & N/A & N/A & N/A & \(-0.46 \pm 0.07\) & \(-0.36 \pm 0.04\) & $-$0.52 $\pm$ 0.05\\
        G311.5$-$0.3 & 4.3 & 3.3 $\pm$ 0.2 & 2,6 & $-$0.5 & \(-0.45 \pm 0.08\) & \(-0.57 \pm 0.07\) & \(-0.52 \pm 0.05\) & $-$0.61 $\pm$ 0.08\\
        G312.5$-$3.0 & 20.8 & 3.5 $\pm$ 0.2 & 16,17 & N/A & \(-0.43\pm0.05\) & \(-0.32 \pm 0.12\) & \(-0.58 \pm 0.35\) & $-$0.10 $\pm$ 0.15\\
        G315.4$-$2.3 & 43.4 & 32 $\pm$ 2 & 12 & $-$0.6 & \(-0.51\pm0.06\) & \(-1.5 \pm 0.4\)& \(-0.58 \pm 0.10\) & $-$0.81 $\pm$ 0.17\\
        G316.3$-$0.0 & 20.9 & 23.1 $\pm$ 0.4 & 2,4,6 & $-$0.4 & \(-0.49\pm0.07\)& \(-0.78 \pm 0.19\) & \(-0.64 \pm 0.05\) & $-$0.78 $\pm$ 0.26\\
        G318.9+0.4 & 19.2 & 6.2 $\pm$ 0.3 & 18 & $-$0.2? & \(-0.24 \pm 0.07\) & \(-0.93 \pm 0.09\)& \(-0.35 \pm 0.05\) & $-$1.2 $\pm$ 0.1\\
        G319.9$-$0.7 & 4.8 & 0.74 $\pm$ 0.04 & 19 & $-$0.26 & \(-0.36\pm 0.04\) & $-$0.36 $\pm$ 0.13 & $-$0.23 $\pm$ 0.13 & $-$0.62 $\pm$ 0.05\\
       G326.3$-$1.8 & 37.9 & 150 $\pm$ 1 & 5,20-22 & varies & \(-0.33 \pm 0.02\) & \(-1.1 \pm 0.1\) & $-$0.27 $\pm$ 0.04 & $-$1.2 $\pm$ 0.1\\
        G327.4+0.4 & 21.6 & 27.7 $\pm$ 0.4 & 5,6,12 & $-$0.6 & \(-0.50\pm0.03\) & \(-1.1 \pm 0.1\) & $-$0.39 $\pm$ 0.21 & $-$0.86 $\pm$ 0.11\\
        G327.4+1.0 & 12.3 & 2.1 $\pm$ 0.1 & 6 & N/A & \(-0.38 \pm 0.09\) & \(-0.61 \pm 0.31\) & $-$0.27 $\pm$ 0.35 & +0.15 $\pm$ 0.31\\
        G328.4+0.2 & 5.6 & 15.6 $\pm$ 0.3 & 2,6,23-25 & 0 & \(-0.11\pm 0.01\) & \(-0.39 \pm 0.09\) & \(-0.17 \pm 0.09\) & $-$0.49 $\pm$ 0.05 \\
        G337.3+1.0 & 12.9 & \(17.0\pm0.4\) & 5,12 & $-$0.55 &  \(-0.43\pm 0.02\) & \(-0.26 \pm 0.14\) & $-$0.49 $\pm$ 0.17 & $-$0.09 $\pm$ 0.09\\
        G340.4+0.4 & 7.6 & 6.95 $\pm$ 0.05 & 6,11,26,27 & $-$0.4 & \(-0.46\pm 0.05\) & \(-0.59 \pm 0.04\) & \(-0.54 \pm 0.03\) & $-$0.61 $\pm$ 0.06\\
        G340.6+0.3 & 5.6 & 5.2 $\pm$ 0.1 & 6,11,26,27 & $-$0.4? & \(-0.38\pm0.11\) & \(-0.54 \pm 0.07\) & \(-0.39 \pm 0.04\) & $-$0.61 $\pm$ 0.04\\
        G341.9$-$0.3 & 7.1 & 3.3 $\pm$ 0.1 & 6,11,27 & $-$0.5 & \(-0.57\pm0.16\) & \(-0.73 \pm 0.09\) & $-$0.52 $\pm$ 0.07 & $-$0.65 $\pm$ 0.05\\
        G344.7$-$0.1 & 9.8 & 3.0 $\pm$ 0.1 & 6,11,28 & $-$0.3? & \(-0.48 \pm 0.04\) & \(-0.82 \pm 0.11\) & $-$0.34 $\pm$ 0.09 & $-$0.74 $\pm$ 0.17\\
        G345.1$-$0.2 & 6.2 & 2.5 $\pm$ 0.1 & 6,29,30 & $-$0.7? & $-$0.38 $\pm$ 0.06 & \(-0.55 \pm 0.16\) & $-$0.26 $\pm$ 0.33 & $-$0.39 $\pm$ 0.05\\
        \hline
    \end{tabular}
    \caption{Known SNRs studied in this paper. The sizes are based on the region of flux integration. The flux densities are the 943 MHz values determined with EMU, some of which are taken from \cite{Ball2025}. The catalogue spectral index values come from the \cite{Green_cat} catalogue. The broadband indices are determined using fluxes obtained from published work, the EMU values provided, and, in some cases, data from the SMGPS and GLEAM. The sub-band, T-T plot, and Taylor term indices are the in-band indices determined here. Flux references used in the determination of broadband indices:
    (1) \cite{Milne1969} (2) \cite{Shaver1970} (3) \cite{Dickel1973} (4) \cite{Milne1975} (5) \cite{Milne1989} (6) \cite{Whiteoak1996} (7) \cite{Filipovic2005} (8) \cite{Lockhart1977} (9) \cite{Braun1986} (10) \cite{Gaensler2003} (11) \cite{Clark1975} (12) \cite{Caswell1975} (13) \cite{Gaensler1998} (14) \cite{Gelfand2013}
    (15) \cite{Reynolds2013} (16) \cite{Duncan1997} (17) \cite{Kane2003}
    (18) \cite{Whiteoak1993} (19) \cite{Ng2010} (20) \cite{Clark1975b} (21) \cite{Milne1979} (22) \cite{Dickel2000} (23) \cite{Caswell1980} (24) \cite{Gaensler2000} (25) \cite{Gelfand2007} (26) \cite{Caswell1983} (27) \cite{Dubner1996} (28) \cite{Dubner1993} (29)\cite{Green2014} (30) \cite{HurleyWalker2019}.
    }
    \label{tab:known_snrs}
\end{table*}

Figure~\ref{fig:snr_indices} compares the three in-band spectral index techniques for the sample of SNRs as functions of size and surface brightness. The gray shaded region indicates the expected range of values for SNR indices, though sources with a pulsar wind nebula (PWN) may have flatter indices that fall above this range. From the plots we can see a divergence of the index values with increasing size and with decreasing surface brightness for the sub-band and Taylor term approaches. The divergent sub-band values are always too steep, falling below the expected value range, while the Taylor term values diverge in both directions. The T-T plot values do not show the same divergence, though the uncertainties do increase with decreasing surface brightness.

\section{Discussion} \label{sec:discuss}

\subsection{Comparing In-band Index Techniques}

\begin{figure*}
    \centering
    \subfigure{\includegraphics[width=0.49\linewidth]{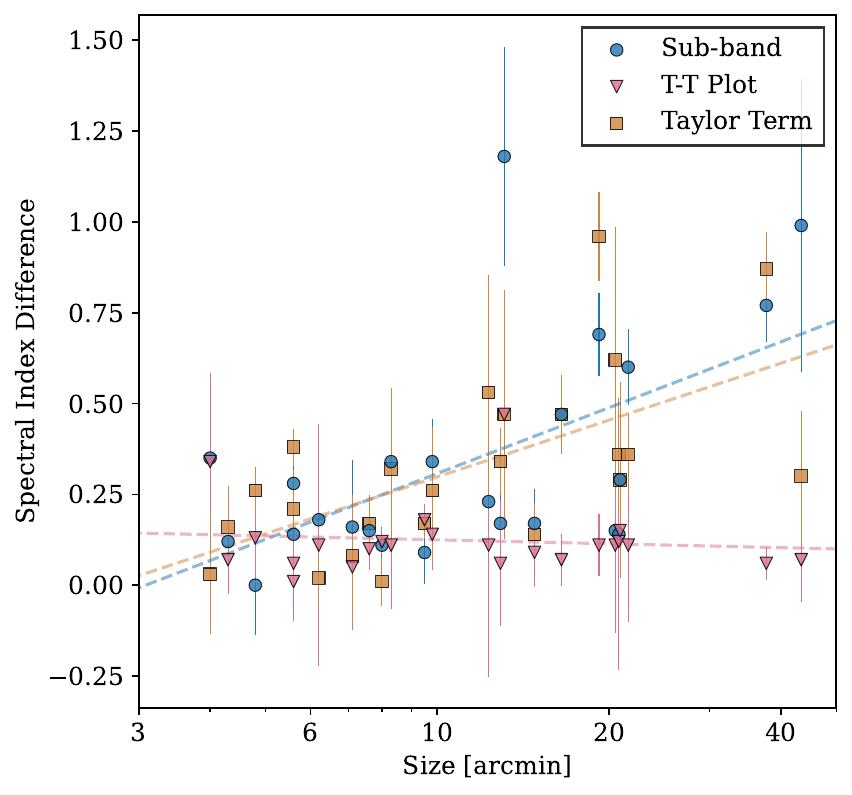}}
    \subfigure{\includegraphics[width=0.49\linewidth]{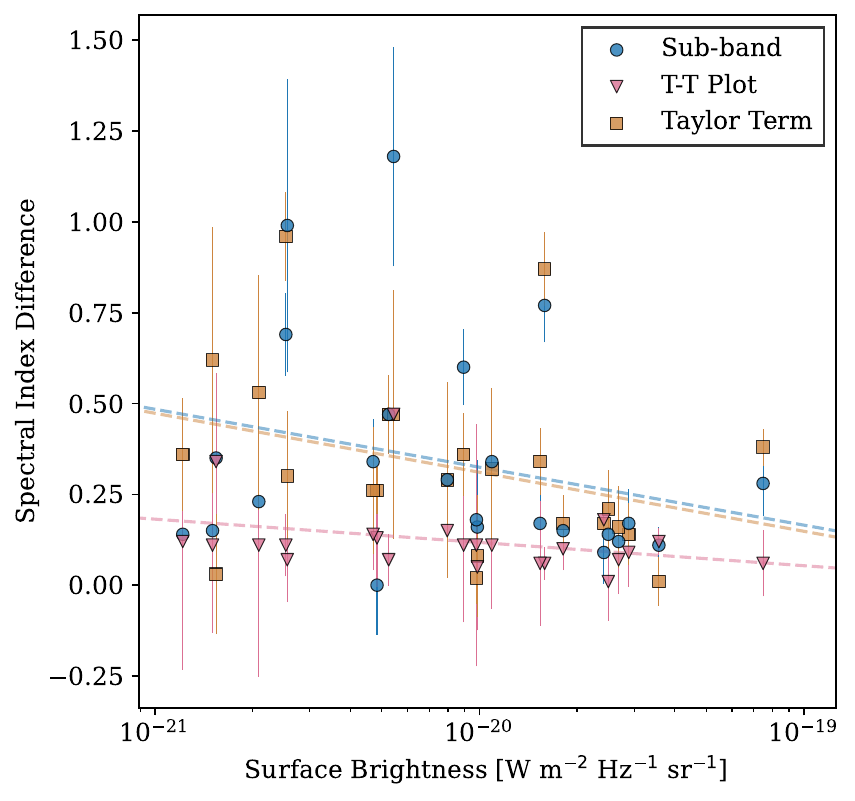}}
    \caption{Comparing the results of the three in-band index techniques with the broadband index values for the sample of SNRs as functions of source size and surface brightness. The spectral index difference is taken as the absolute value of the difference between the in-band and broadband indices. The dashed lines represent the fit lines corresponding to each of the techniques. The sub-band and Taylor term indices diverge from the expected values as source size increases and as surface brightness decreases. The T-T plot method performs consistently across the range of sizes and surface brightnesses studied.}
    \label{fig:index_compare}
\end{figure*}

The goal of this paper is to evaluate the reliability of three in-band spectral index techniques. In Figure~\ref{fig:index_compare}, we compare the performance of the in-band techniques for the sample of SNRs by taking the difference between the broadband index value, determined using published data to provide a point of comparison, and the index values from each of the in-band techniques. The absolute value of the difference is graphed as functions of size and surface brightness. The error bars were determined based on the uncertainties in the broadband and in-band indices. The dashed lines are the fit lines corresponding to each technique. 

Here we see the sub-band and Taylor term method fit lines are very similar in both graphs, showing a significant divergence from the broadband values as the sources become larger or fainter. The T-T plot method performs consistently across the size and surface brightness ranges probed. The fit line shows that, on average, the T-T plot indices fall within 0.2 of the broadband indices across all source types. It should be noted that this analysis assumes the broadband values are reliable. This is not an unreasonable assumption as all of the broadband indices used were calculated with a minimum of five data points and are consistent with the expected index values for SNRs. Additionally, when we compare the in-band indices to the theoretical value range for spectral indices of SNRs (see Figure~\ref{fig:snr_indices}), we find similar results. 

\subsubsection{Sub-band Integration}

The sub-band integration method works reasonably well for small, bright sources but starts to become less trustworthy as the source sizes become larger and the emission becomes fainter and more diffuse. Above a source size of around 10$'$, we start to see noticeable deviations from the expected value range in most of the studied SNRs. ASKAP is sensitive to extended emission at maximum scales of 43$'$ to 60$'$ across the EMU band though sensitivity starts to decline at scales above a few arcminutes \citep{Hopkins2025}. Hence, with the sub-band method we start to see noticeable divergences from the expected value range at source sizes of around 20\% of this maximum scale, which could reasonably be attributed, at least in part, to the decrease in sensitivity. It should be noted, however, that the morphology of the source is a relevant factor. The sub-band indices may be reliable for sources larger than 10$'$ if the emission is dominated by small-scale features that are well recovered by the interferometer. Because the divergent values are generally too steep, sub-band indices may be useful as lower limits.

The EMU fields have a median rms noise level of around 30 $\mu$Jy beam$^{-1}$ with 95\% of observed fields having a noise level $<$45 $\mu$Jy beam$^{-1}$ \citep{Hopkins2025}. Notably, Galactic plane fields will typically have higher noise levels than high-latitude tiles. We find the reliability of the sub-band indices starts to decline at surface brightnesses that correspond to around 50$-$100 times the median noise level. The exact relation between source brightness and noise will depend on the environment of that particular source, specifically the S/N, which may explain why we see some deviations from this trend. 

We also use the four sub-band images to construct spectral index maps, which can help to reveal changes in index across the source. In some cases, it may be possible to obtain a spectral index for part of a source, perhaps for a PWN or bright shell feature. If there are problems with the calculated indices, the index maps can help to reveal if these issues only affect part of the source. Typically, it is obvious when the sub-band technique is not reliable as this method tends to produce indices that are implausibly steep, diverging significantly from the expected range of index values. The index maps reveal that this steepening often corresponds to the larger-scale, more diffuse parts of the source. When noise is a factor, this may also be evidenced from the index maps as we can see small-scale fluctuations in the index at low S/N that do not appear to correspond to real changes across the source. 

\subsubsection{T-T Plots}

The T-T plot method performs the best overall, particularly for large sources that can be assumed to have significant missing flux. When the missing emission is constant across a region, it behaves as an offset that affects the intercepts of the T-T plots but not the slopes. Because of this, T-T plots are particularly well suited to studying extended sources with interferometers. While we do see an increase in uncertainties in the T-T plot indices with decreasing surface brightness, we find that this method produces reliable results across the size and surface brightness ranges studied. 

In some cases, multiple components can be seen within the T-T plots, particularly in composite sources where we see a flatter PWN component that diverges from the steeper shell component. In other cases, we find a change in spectral index between the brighter and fainter parts of the source, such as in the \ion{H}{2} regions where we find bright cores with positive indices surrounded by flatter optically thin emission. In these cases, a piecewise linear regression may be appropriate if we are interested in determining the indices of the different components rather than an average for the entire source. For composite SNRs, we may also produce multiple sets of T-T plots for the different regions of interest, separating the PWN and shell components. 

The T-T plot index technique has considerable potential working with new radio surveys as the higher resolution provides more data points. The main disadvantage to the T-T plot method is \rev{we lose spatial information and cannot} produce spectral index maps. However, because the in-band T-T plot method requires the production of sub-band images, the sub-band and T-T plot techniques may be used synergistically as they require the same data products. Broadband T-T plots may also perform very well, especially using two different bands from the same telescope, as a wider frequency range would reduce uncertainties, particularly for fainter sources. 

\subsubsection{MFS Taylor Terms}

The Taylor term index technique may also be reliable for small, bright sources and can be used to produce spectral index maps. However, we find this technique to be the least dependable overall. While it does seem to produce reliable results for SNRs smaller than around 10$'$, we see the \ion{H}{2} region indices start to diverge from the expected value range above sizes of around 2$'$. This divergence results in overlap between the populations of SNRs and \ion{H}{2} regions and means that this technique cannot reliably distinguish between sources of thermal and nonthermal emission above approximately 2$'$. Similarly, this method may work for very bright sources, but we start to see significant overlap between the indices of the studied SNRs and \ion{H}{2} regions with decreasing surface brightness.

With the Taylor term approach, we did not account for background variations, which could help to explain why we see issues start to arise with \ion{H}{2} regions before they arise in SNRs. \ion{H}{2} regions have less well-defined boundaries and are often found in dense regions with higher noise. The Taylor term uncertainties provided here are based on the noise level of the surrounding region and thus are a reflection of the S/N for a particular source. Particularly with the HII regions, we see that many of the divergent sources have large error bars, indicating that high-noise regions may be particularly problematic.

The main advantage of the Taylor term method is efficiency. Because this technique does not require any additional data processing, it is currently the quickest and easiest to implement. However this advantage is not inherent as modifications could be made to the ASKAP pipeline to include the production of sub-band images. In general, the Taylor term method may be useful for quick tests, particularly for small, bright objects, but should be assumed to have limited reliability for extended sources. 

It must be noted that this analysis does not consider the systematic uncertainties in the Taylor term indices that may result from the imaging process itself. Imperfect deconvolution can produce correlated errors between the Taylor terms \citep{Rau2011}. \rev{In our work, we rely solely on the Taylor term maps produced through the ASKAP pipeline. It may be possible to improve the performance of the Taylor term method by optimizing the cleaning process for the source of interest. Algorithms optimized for spectral index measurements \citep{Junklewitz2015,Hsieh2026}, faint extended emission in particular, may see better results than those presented here.}

\subsection{Distinguishing Thermal and Nonthermal Sources}

\begin{figure*}
    \centering
    \subfigure{\includegraphics[width=\linewidth]{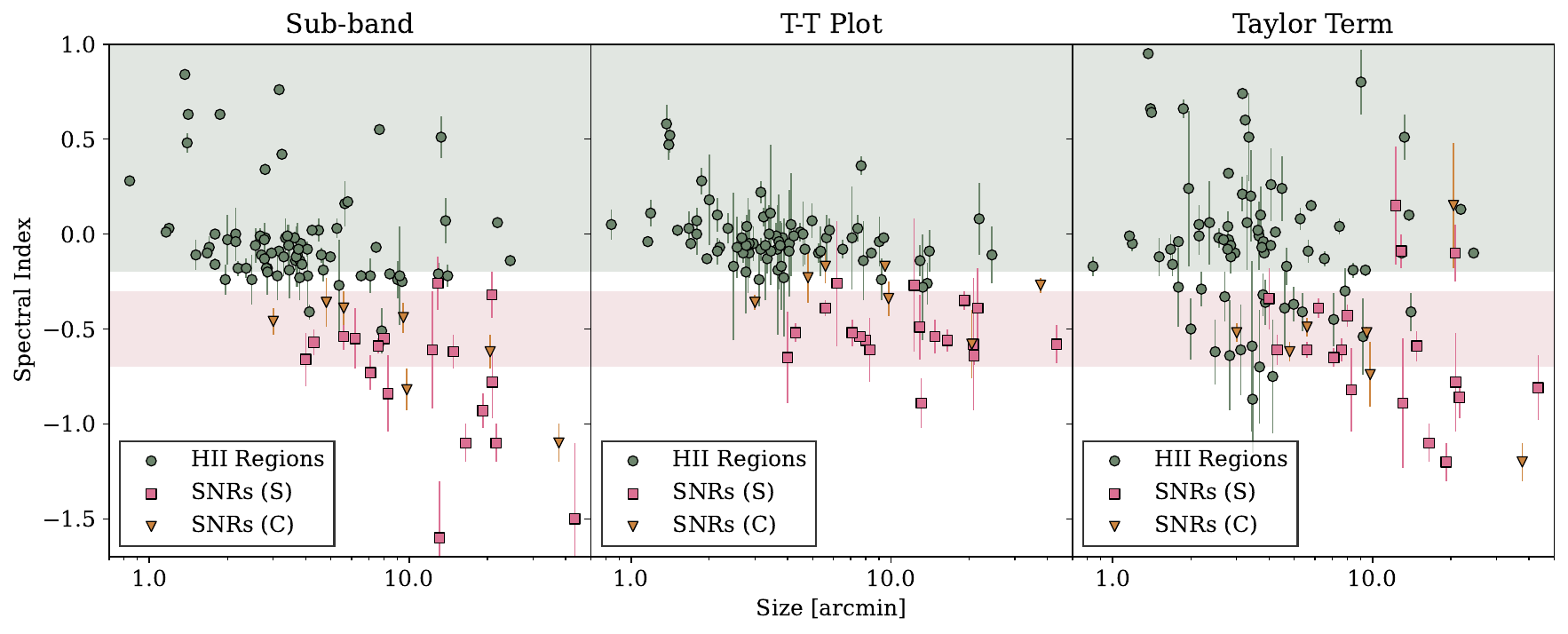}}
    \subfigure{\includegraphics[width=\linewidth]{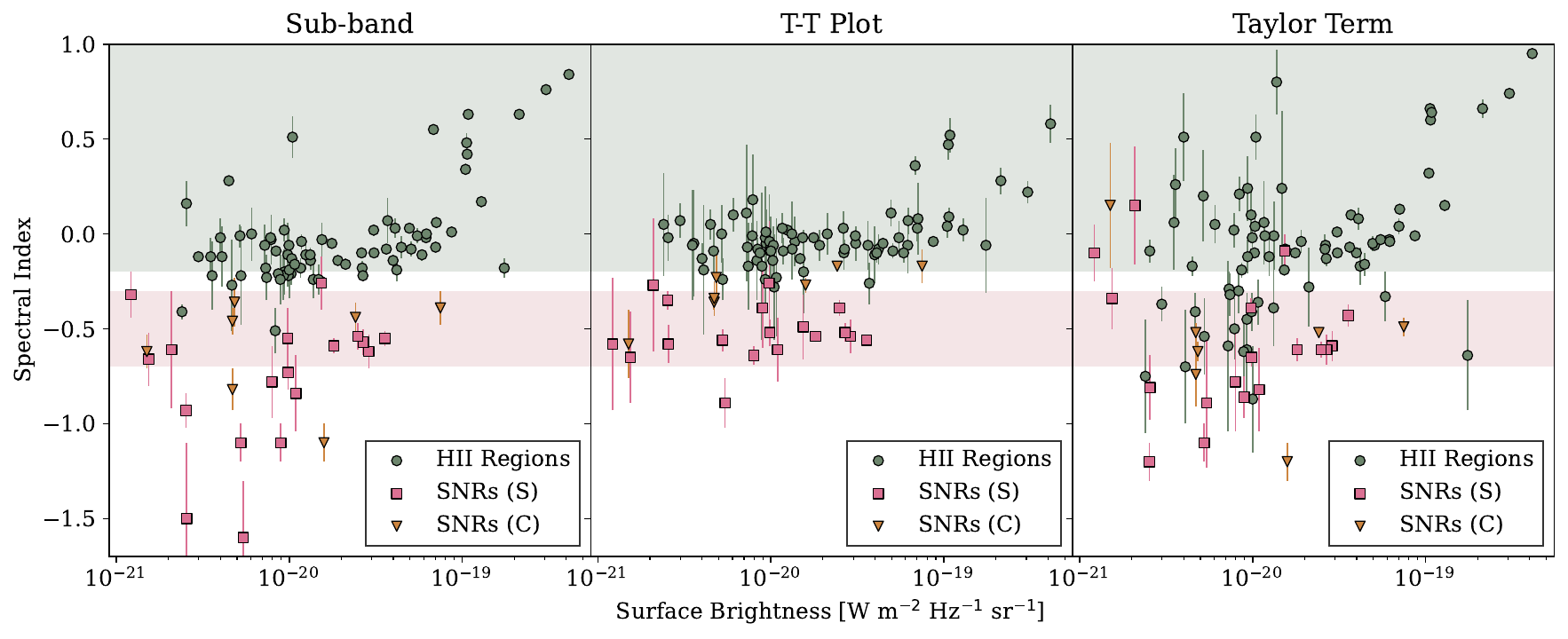}}
    \caption{Comparing in-band spectral index techniques for the samples of \ion{H}{2} regions and SNRs (S=shell-type and C=composite/centre-filled) as functions of size and surface brightness. The shaded regions indicate the expected value ranges for the two source types. With the sub-band method, we see the index values steepen significantly with increasing size and with decreasing surface brightness, particularly for the SNR population. We do not see this same steepening with the T-T plot method, though we do see an increase in the index uncertainties as the sources become fainter. In the Taylor term panels, we see significant overlap between the two populations, an effect that worsens with increasing size and decreasing surface brightness.}
    \label{fig:comparison}
\end{figure*}

Figure~\ref{fig:comparison} compares the performance of the in-band index techniques for the samples of SNRs and \ion{H}{2} regions as functions of size and surface brightness. We are interested in the size and surface brightness range where the index values for the two populations start to overlap. This sets a limit for when we can trust the spectral index method to reliably distinguish between thermal and nonthermal emission. The shaded regions are the expected ranges of values. SNRs are further distinguished by type, separating the shell-type (S) SNRs, which are expected to have steeper indices, from the composite and centre-filled type (C), which can have flatter indices that fall above the shaded SNR range. 

These graphs clearly demonstrate the superior performance of the T-T plot method across the studied parameter space. The sub-band indices deviate more significantly from the expected values with increasing size and decreasing surface brightness, particularly for the sample of SNRs. We see that the small faint sources can be distinguished using both techniques but particularly with the sub-band method, the populations start to overlap as the sources get larger and, even more so, as they get fainter. The Taylor term indices show significant overlap between the two populations and thus they cannot be used to reliably distinguish between extended sources of thermal and nonthermal emission. It is also notable that while the sub-band method shows a steepening effect toward negative values as source size increases and brightness decreases, the Taylor term indices for the SNRs also diverge toward positive values, meaning the indices cannot even be used as lower limits. 

\subsection{\rev{Potential Sources of Spectral Steepening}}

The results presented here show that a spectral steepening toward negative values is often observed as sources become larger and fainter, particularly evident with the sub-band and Taylor term methods. \rev{One potential cause of this is missing short spacings.} Because missing flux as a consequence of missing short spacings is a frequency-dependent effect, the higher-frequency channels will be impacted more than the lower ones. Thus, the high-frequency channels will be missing more flux, resulting in spectral steepening. \rev{This could be tested} by using single-dish observations to fill in the missing flux. The ongoing survey PEGASUS (POSSUM EMU GMIMS All-Stokes UWL Survey), which is being conducted with the Parkes telescope and aims to cover the entire southern sky \citep{Hopkins2025,Gaensler2025}, will be useful in this endeavor. 

\rev{Frequency-dependent primary beam effects can also lead to spectral steepening \citep{Bhatnagar2013}. While primary beam correction has been applied to the EMU data as part of the ASKAP pipeline, residual primary beam effects have not been ruled out as a potential source of the observed deviations.}
\rev{Follow-up work is required to determine the extent to which differences between methods can be attributed to missing short spacings, primary beam effects, deconvolution uncertainty, or background fluctuations.}


\subsection{Sub-band Integration: Optimizing Sub-band Choice} \label{sec:subband_options}

In this work, we chose to calculate our sub-band indices by re-imaging the EMU data to produce four 72 MHz wide sub-band images. This is not the only way to apply the sub-band integration technique. Alternatively, we could have chosen to use the POSSUM Stokes I data cubes, which have 288 1 MHz channels. This approach has the advantage of not requiring additional data processing but the S/N for each channel will be much lower, as the image noise scales as $1/\sqrt{\Delta\nu}$, where $\Delta\nu$ is the channel width \citep{Thompson2017}. This means the 1 MHz channels will have approximately 8.5 times higher noise levels than the 72 MHz sub-bands. The choice to re-image the data was made with the aim of applying this technique to fainter sources. Here, we briefly explore the effectiveness of this approach through comparison to the cube data. 

A limited analysis finds that for bright sources, using the 288 channels is comparable to using four and both methods produce similarly accurate results. However, the higher noise levels in the 1 MHz channels mean the indices become unreliable more quickly as source brightness decreases, as expected given the relation between index accuracy and S/N. For fainter sources, we find the data cube spectra are dominated by noise and appear as increasingly nonlinear with lower S/N. We also find that the indices determined using 1 MHz channels can be significantly impacted by the choice of background, particularly for the fainter sources. This sensitivity to background variations results in larger index uncertainties.

We also explored the possibility of producing the four 72 MHz wide sub-bands through averaging the 1 MHz channels, rather than through re-imaging. We find that this approach seems to produce results that are effectively identical to those produced using the 1 MHz wide channels for the bright sources where the resultant indices are reliable. For the lower-S/N sources, the averaging can help to smooth out the excessive noise and create a more linear spectrum. However, the sensitivity to the choice of background seems to persist through the averaging process and results in larger uncertainties for faint sources.

In general, this rudimentary analysis suggests that using the 1 MHz channels is sufficient for bright sources. For lower-S/N sources, the re-imaged data produce better results, particularly for smaller sources where the majority of the flux is properly recovered. The optimal choice of sub-band width will ultimately depend on the source type and study goals. The main disadvantage to re-imaging is the additional work required but the results produced are either equivalent or superior to those obtained from the data cubes. Notably, the sub-band images are of additional value as they can be used to produce T-T plots.

\section{Conclusions} \label{sec:conc}

In this work, we evaluated the performance of three different in-band spectral index techniques for extended sources of radio emission, primarily \ion{H}{2} regions and SNRs. The techniques studied are sub-band integration, T-T plots, and using the ratio of MFS Taylor terms. The spectral indices produced with the in-band techniques are compared to the theoretical values and to broadband indices determined using flux values from previous studies. 

We find the T-T plot method to be the most reliable, producing consistent results across the range of source sizes and brightnesses studied for both SNRs and \ion{H}{2} regions. While the sizes of the uncertainties increase with decreasing surface brightness, the index values largely remain consistent with both the theoretical value range and the broadband indices. The sub-band integration technique produces reliable results for small and bright sources but deviates from the expected value range with increasing source size and decreasing source brightness. The Taylor term method deviates the most significantly from the expected value range and cannot reliably distinguish between sources of thermal and nonthermal emission for sources larger than a couple of arcminutes. These findings are significant as Taylor terms are commonly used to determine in-band indices. \rev{Further research is required to better understand the primary drivers of the deviations seen with the sub-band and Taylor term methods and how these issues might be ameliorated.}

Because the T-T plot technique and the sub-band integration technique can be applied using the same data products, the best approach may be to use both methods. The T-T plot method produces the best integrated index values, though \rev{spatial information is lost as the solution is integrated over all pixels. The sub-band images are useful because they can be used to produce index maps, which provides insight into spectral changes across the source and can highlight potential problems.} If there is agreement between the T-T plot indices and the sub-band integration indices, this provides further validation. If there are discrepancies between these two values, the spectral index maps can reveal whether they may be related to spectral changes across the source or if they are a consequence of instrumental effects.

Ultimately, due to the small frequency range, in-band spectral indices will have larger uncertainties than broadband values and these techniques should be applied with caution and nuance, particularly when dealing with extended sources. \rev{We have shown that there is sufficient information in the EMU data to produce reliable in-band spectral indices, but that recovering this information on a pixel-by-pixel basis remains challenging for extended, low surface brightness emission. Because spectral index accuracies of at least 0.1 are required to properly distinguish thermal and nonthermal sources of emission, in-band index algorithms that can achieve this level of precision for low surface brightness sources while preserving spatial information are highly desirable.}

\begin{acknowledgments}
This scientific work uses data obtained from Inyarrimanha Ilgari Bundara, the CSIRO Murchison Radio-astronomy Observatory. We acknowledge the Wajarri Yamaji People as the Traditional Owners and native title holders of the Observatory site. CSIRO’s ASKAP radio telescope is part of the Australia Telescope National Facility (https://ror.org/05qajvd42). Operation of ASKAP is funded by the Australian Government with support from the National Collaborative Research Infrastructure Strategy. ASKAP uses the resources of the Pawsey Supercomputing Research Centre. Establishment of ASKAP, Inyarrimanha Ilgari Bundara, the CSIRO Murchison Radio-astronomy Observatory and the Pawsey Supercomputing Research Centre are initiatives of the Australian Government, with support from the Government of Western Australia and the Science and Industry Endowment Fund. 

The authors acknowledge the use of the Canadian Advanced Network for Astronomy Research (CANFAR) Science Platform operated by the Canadian Astronomy Data Center (CADC) and the Digital Research Alliance of Canada, with support from the National Research Council of Canada (NRC), the Canadian Space Agency (CSA), CANARIE, and the Canadian Foundation for Innovation (CFI). B.D.B., R.K., and E.R. acknowledge the support of the Natural Sciences and Engineering Research Council of Canada (NSERC), funding reference number RGPIN-2022-03499. \rev{We are grateful to an anonymous referee whose comments improved the quality of the paper.}
\end{acknowledgments}

\bibliography{indices}{}
\bibliographystyle{aasjournalv7}

\appendix

\section{Detailed descriptions of selected SNRs} \label{app:a}

\begin{figure*}
    \centering
    \subfigure[G290.1$-$0.8]{\includegraphics[width=0.49\linewidth]{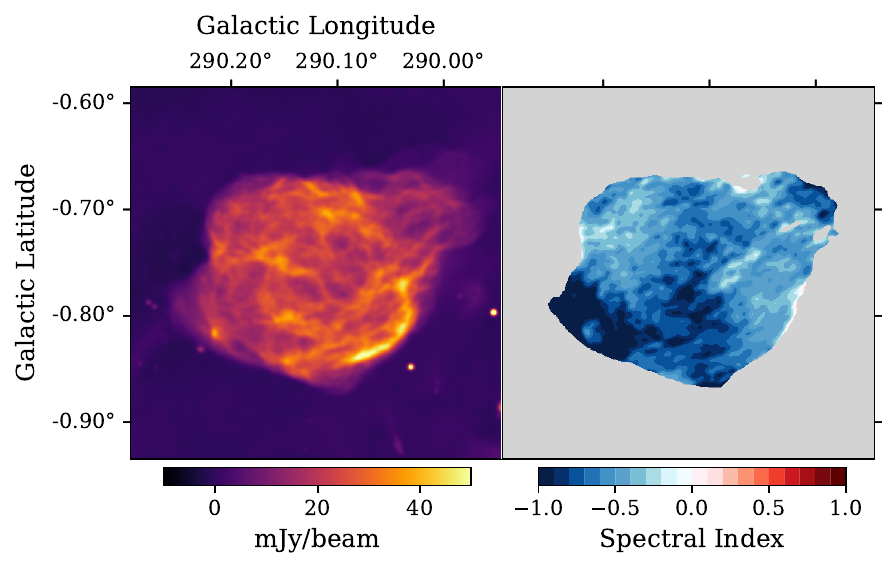}\label{fig:G290.1-0.8}}
    \subfigure[G292.0+1.8]{\includegraphics[width=0.49\linewidth]{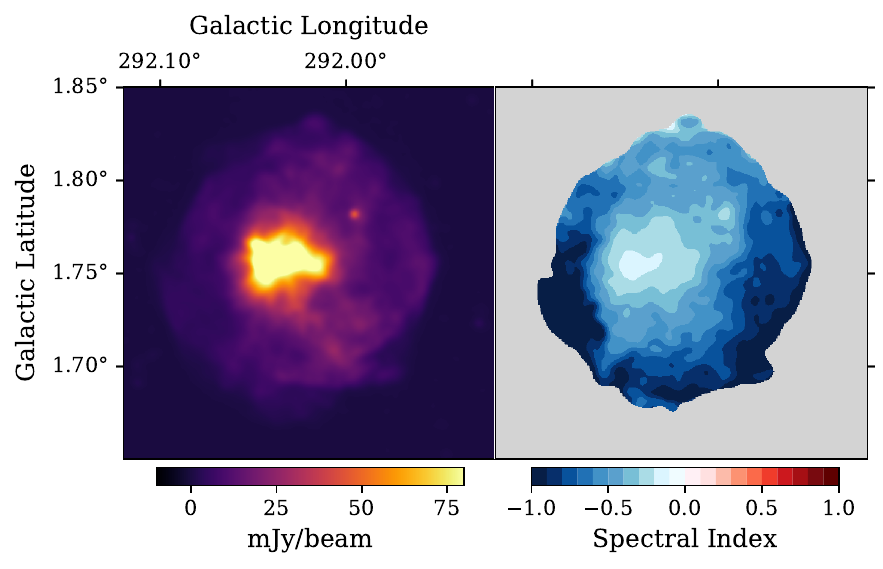}\label{fig:G292.0+1.8}}
    \subfigure[G304.6+0.1]{\includegraphics[width=0.49\linewidth]{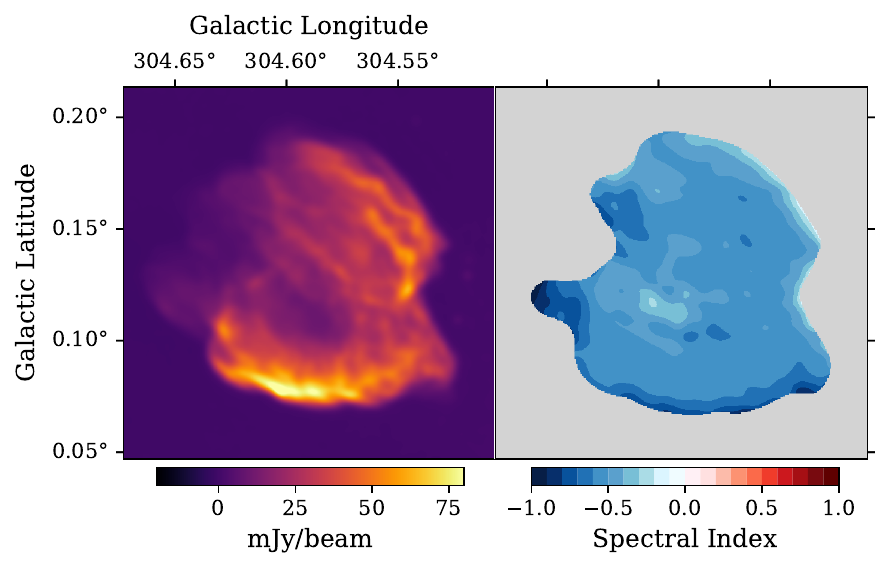}\label{fig:G304.6+0.1}}
    \subfigure[G310.6$-$1.6]{\includegraphics[width=0.49\linewidth]{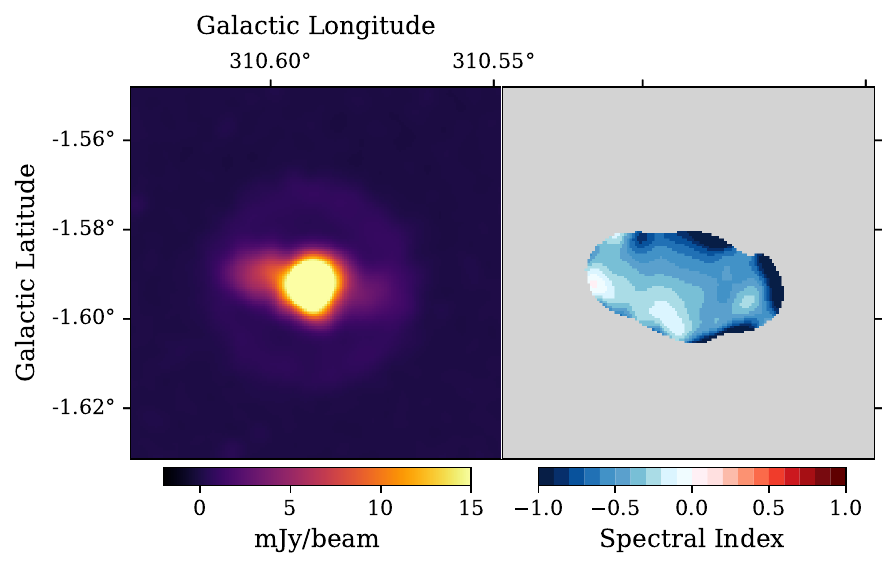}\label{fig:G310.6-1.6}}
    \subfigure[G311.5$-$0.3]{\includegraphics[width=0.49\linewidth]{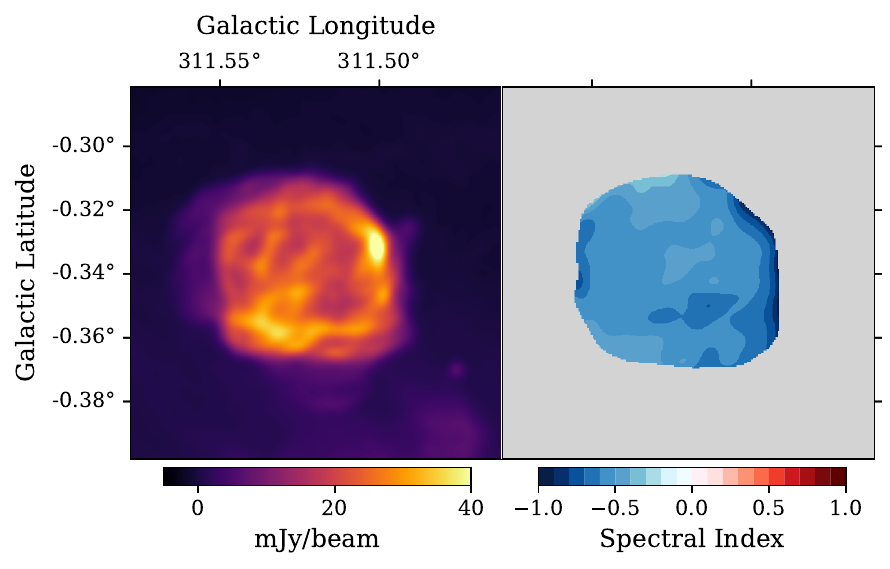}\label{fig:G311.5-0.3}}
    \subfigure[G319.9$-$0.7]{\includegraphics[width=0.49\linewidth]{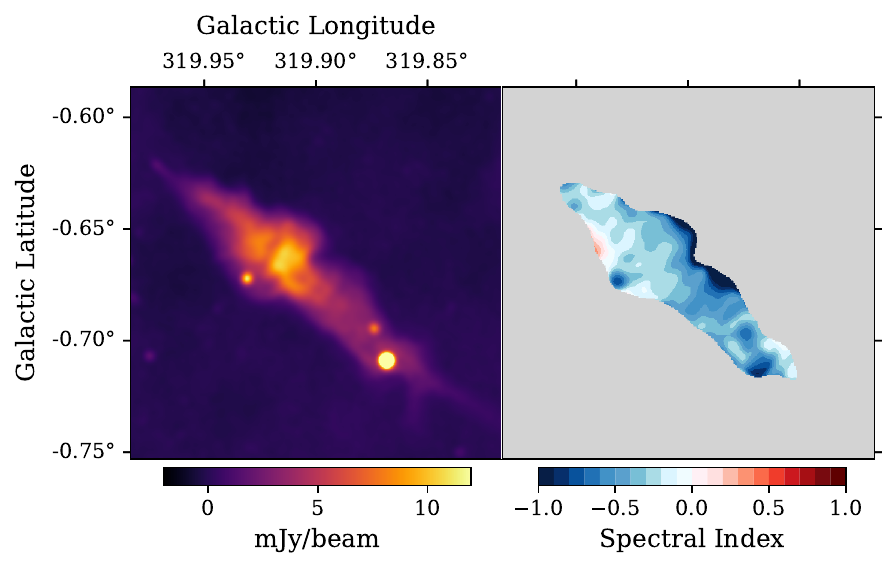}\label{fig:G319.9-0.7}}
    \caption{In-band spectral index maps for known SNRs. The left panels show the 943 MHz Stokes I images from EMU. The right panels show the spectral index maps produced using the four 72 MHz wide sub-bands.}
    \label{fig:snrs_maps}
\end{figure*}

\begin{figure*}
    \subfigure[T-T plot for the shell-type SNR G290.1$-$0.8.]{\includegraphics[width=0.48\linewidth]{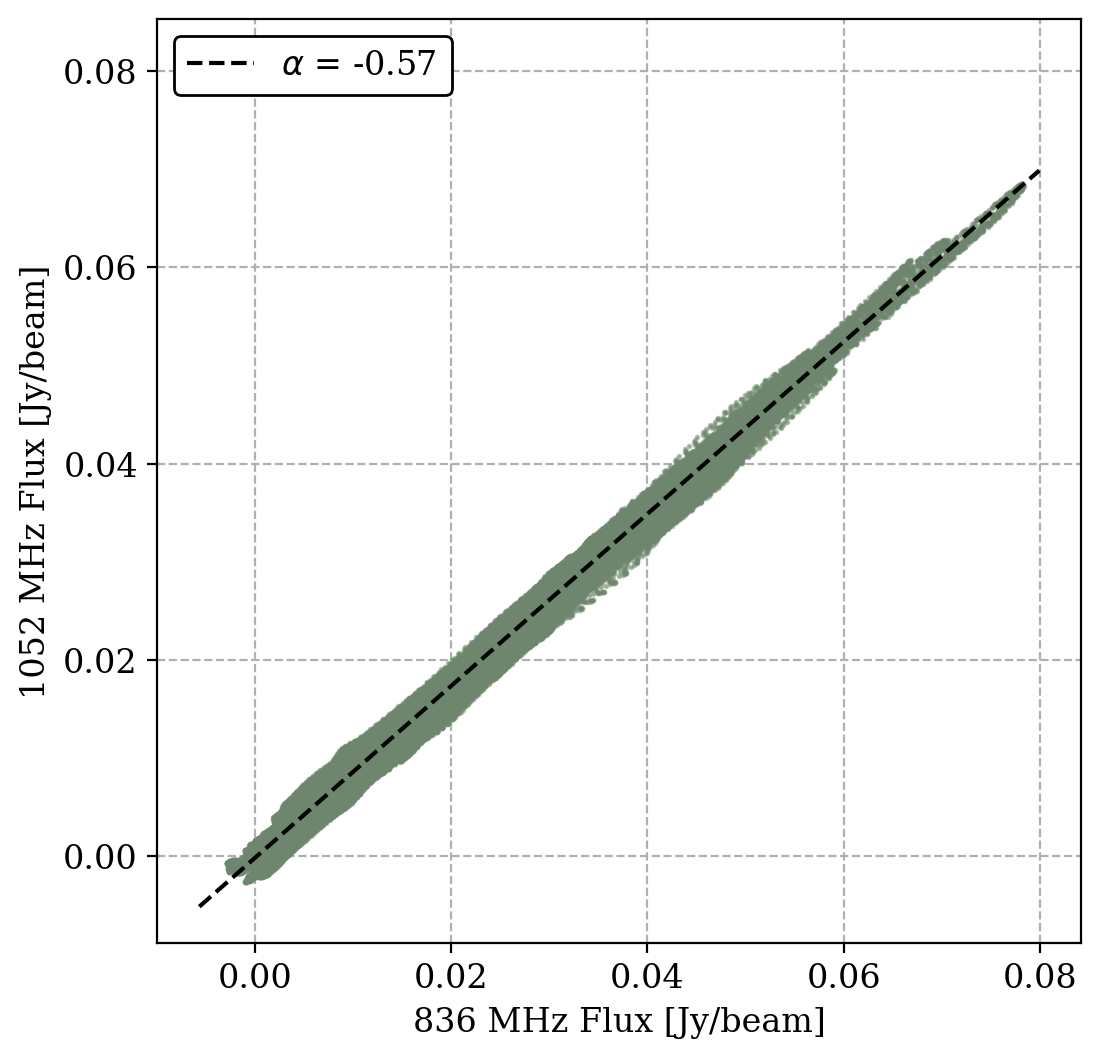}\label{fig:G290.1-0.8_tt}}\hfill
    \subfigure[T-T plot for the shell-type SNR G304.6+0.1.]{\includegraphics[width=0.48\linewidth]{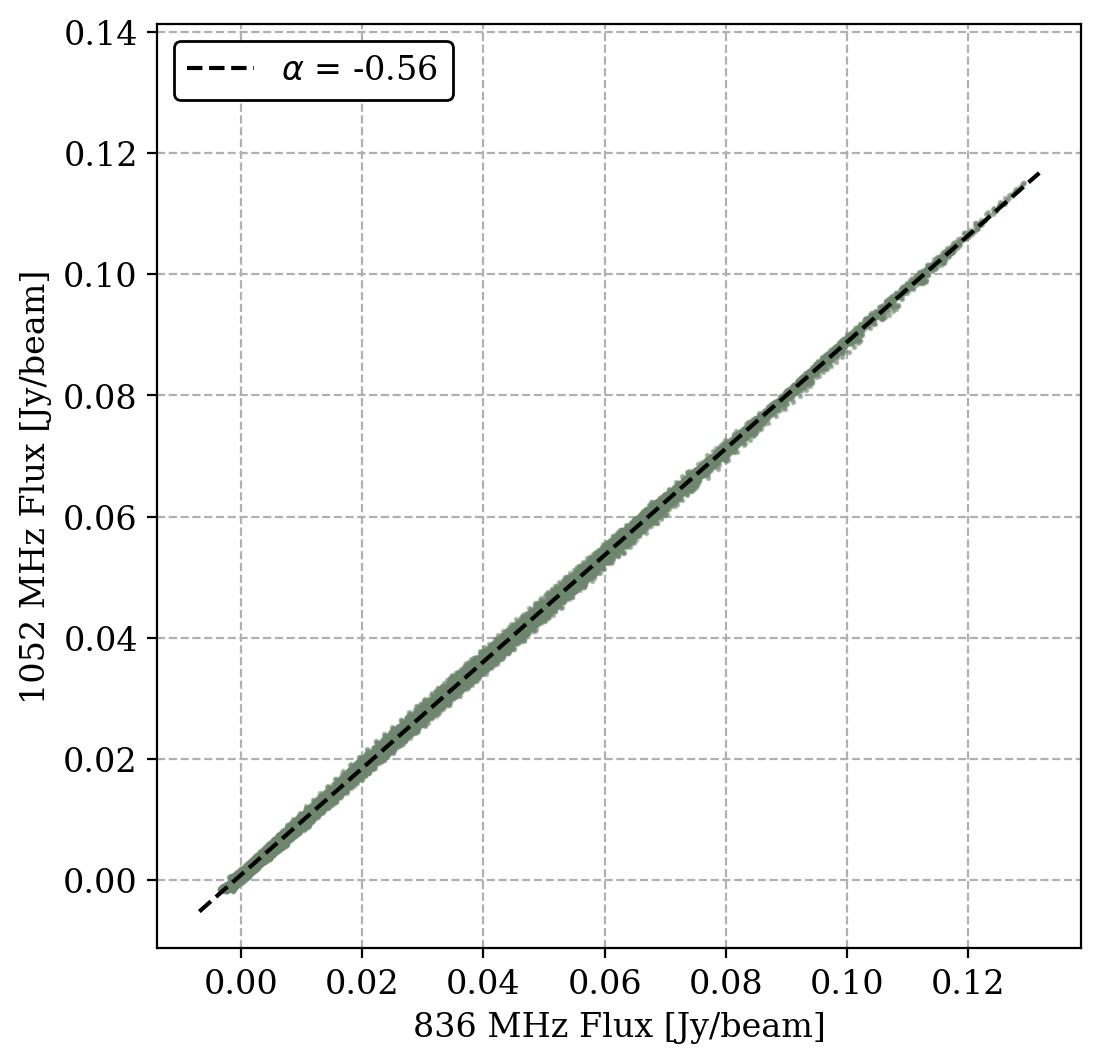}\label{fig:G304.6+0.1_tt}}
    \subfigure[T-T plot for the composite SNR G326.3$-$1.8. We can see multiple components that diverge toward the higher flux values. The steeper component corresponds to the shell while the flatter component corresponds to the PWN.]{\includegraphics[width=0.48\linewidth]{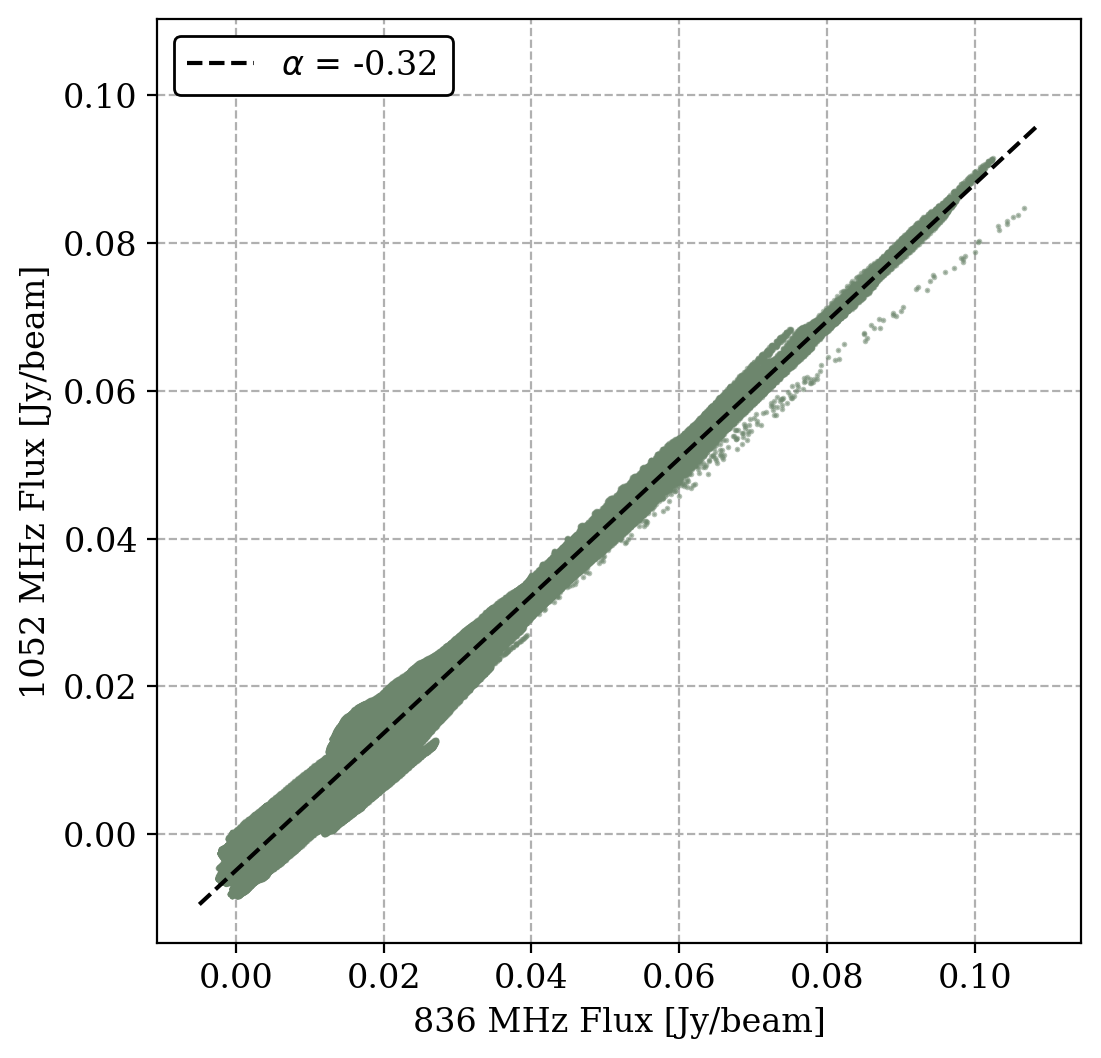}\label{fig:G326.3-1.8_tt}}\hfill
    \subfigure[T-T plot for the center-filled SNR G328.4+0.2. From the T-T plot we can observe the same relatively steeper index in the fainter emission that we see in the index map as the low flux data points deviate slightly from the trend line.]{\includegraphics[width=0.48\linewidth]{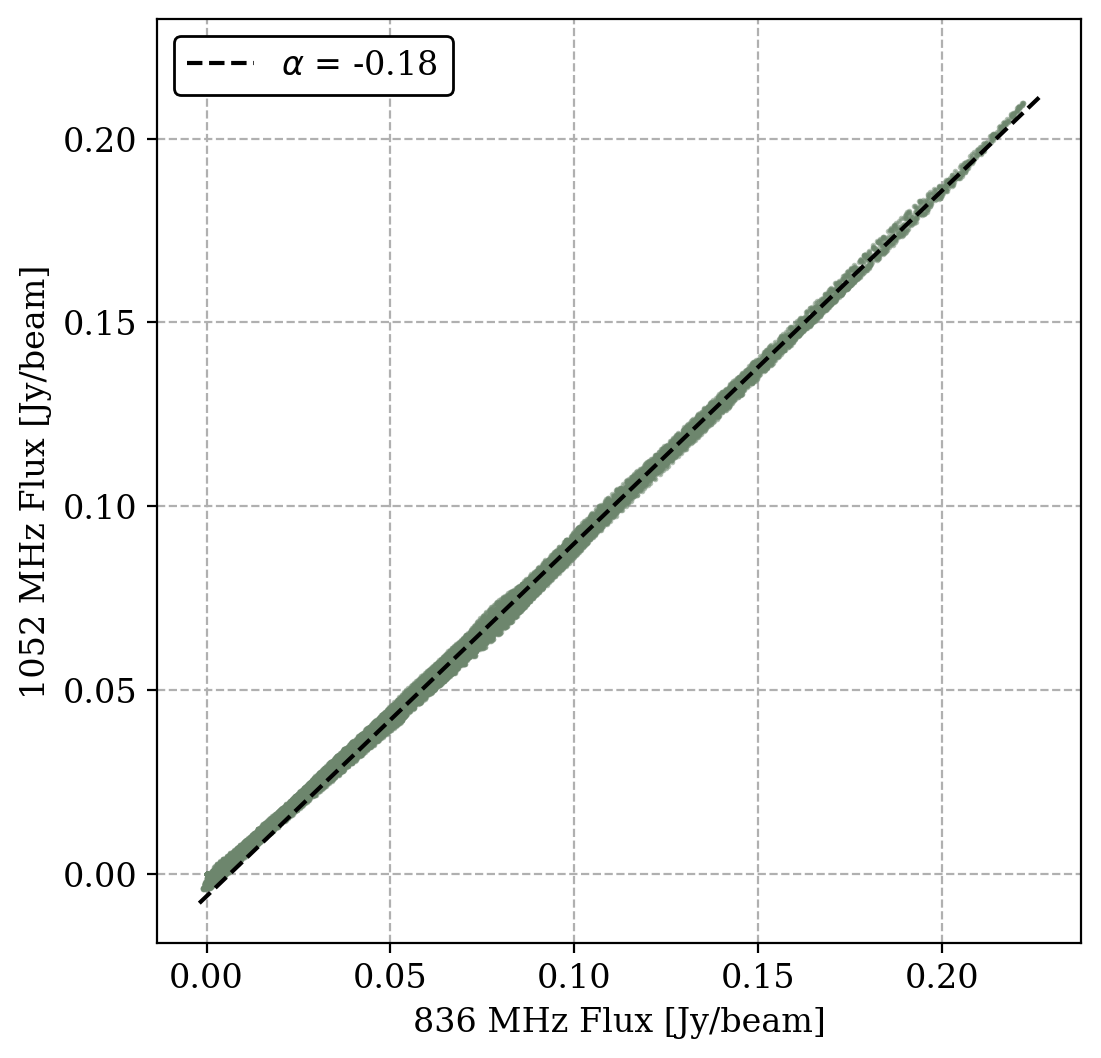}\label{fig:G328.4+0.2_tt}}
    \caption{Sample T-T plots for known SNRs created using the lowest- and highest-frequency sub-bands.}
\end{figure*}

\paragraph{G290.1$-$0.8} 
This source is a relatively large, bright shell-type SNR with many small-scale filamentary structures across its interior. Changes in spectral index across the source can be seen in the sub-band index map (Figure~\ref{fig:G290.1-0.8}). The small-scale bright filaments are well recovered but the fainter diffuse component is likely missing flux, steepening the index. All of the in-band techniques produce values that are slightly steeper than the broadband index, perhaps related to this missing flux. While the in-band indices are steeper than expected for this source, they are still all consistent with the range of expected values for an SNR. The T-T plot index is in closest agreement with the broadband value. We also show a sample T-T plot for this source in Figure~\ref{fig:G290.1-0.8_tt}. The wider spread of data points in the T-T plot for this sources indicates that either it is more affected by noise than the other sources or possibly that there are variations in the spectral index across the source.

\paragraph{G292.0+1.8}
This is a composite SNR with a faint visible shell and a bright central PWN that dominates the majority of the flux. In the spectral index map (Figure~\ref{fig:G292.0+1.8}), we can clearly see a distinction between the flatter index of the PWN and the steeper index of the shell, with the faintest parts of the shell being implausibly steep, indicating there is likely missing flux. We observe similar spectral steepening toward the edges in many other sources that tends to correspond with the fainter, diffuse emission. For this source, the index from the sub-band method is the most similar to the broadband value, though it is slightly steeper. The T-T plot index is flatter, likely because it is more strongly weighted toward the index of the bright central component. The Taylor term index is steeper than the broadband value and the other in-band values.

\paragraph{G304.6+0.1}
This source is an intermediate-sized, bright, shell-type SNR. The index map is relatively smooth, showing little change in the spectrum across the source  (Figure~\ref{fig:G304.6+0.1}). The sub-band and T-T plot indices are slightly steeper than the broadband value, though they are both still consistent with the expected value for a shell-type SNR. The Taylor term value is consistent with the broadband value. We also show a sample T-T plot for this source in Figure~\ref{fig:G304.6+0.1_tt} that shows the same consistency in the index across the source.

\paragraph{G310.6$-$1.6} 
This source is a very small composite SNR. The central PWN is much brighter than the surrounding shell, which is incredibly faint and was only recently discovered with ASKAP \citep{Jing2025}. The shell is too faint to obtain a spectral index. The PWN has a flat negative index, consistent with the expected value for this type of source. The map shows the index is flattest in the lower left edge of the PWN region, steepening toward the upper right (Figure~\ref{fig:G310.6-1.6}). There are not enough published data for this source to obtain a broadband index as it is a small high-latitude source and thus has not been detected in many radio surveys. The three in-band techniques produce different results, all of which are plausible for a PWN, though the T-T plot index is the closest to the typical value range.

\begin{figure*}
    \centering
    \subfigure[G326.3$-$1.8]{\includegraphics[width=0.49\linewidth]{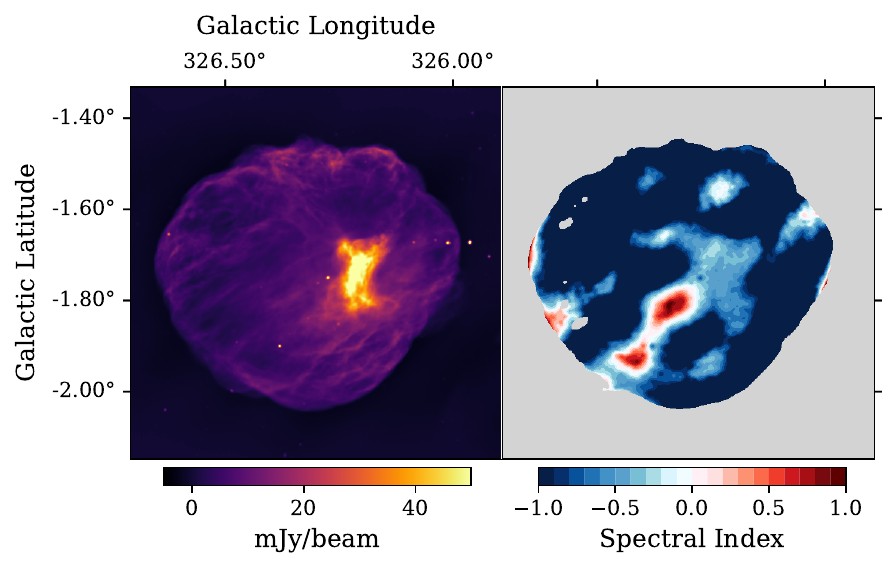}\label{fig:G326.3-1.8}}
    \subfigure[G328.4+0.2]{\includegraphics[width=0.49\linewidth]{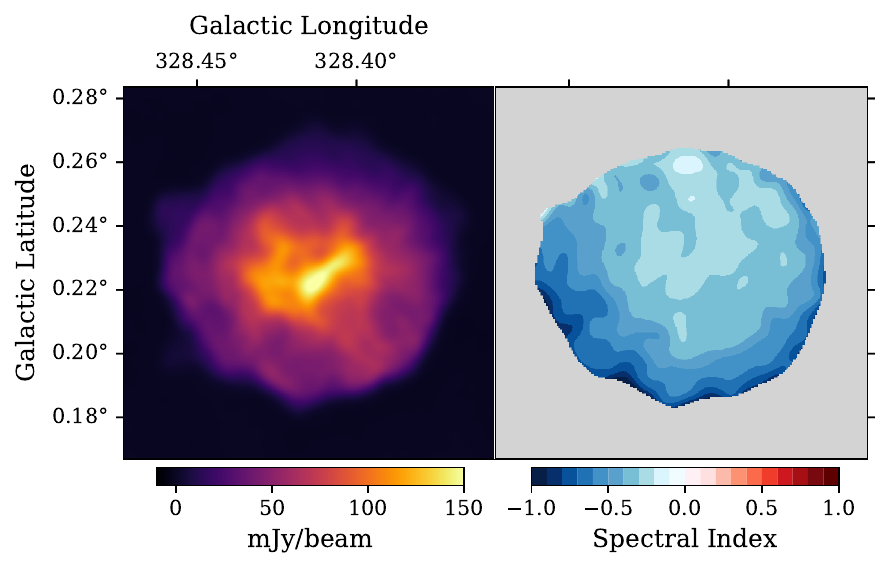}\label{fig:G328.4+0.2}}
    \subfigure[G340.4+0.4]{\includegraphics[width=0.49\linewidth]{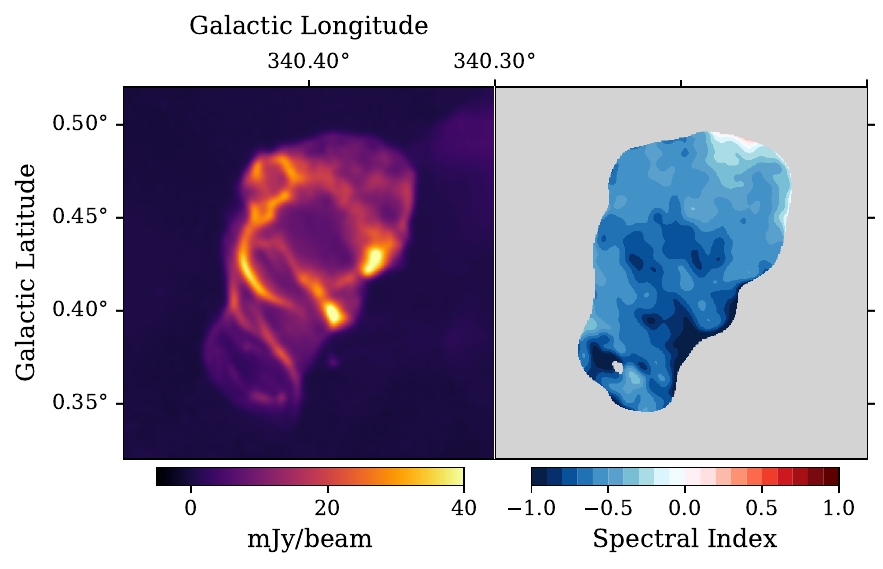}\label{fig:G340.4+0.4}}
    \subfigure[G340.6+0.3]{\includegraphics[width=0.49\linewidth]{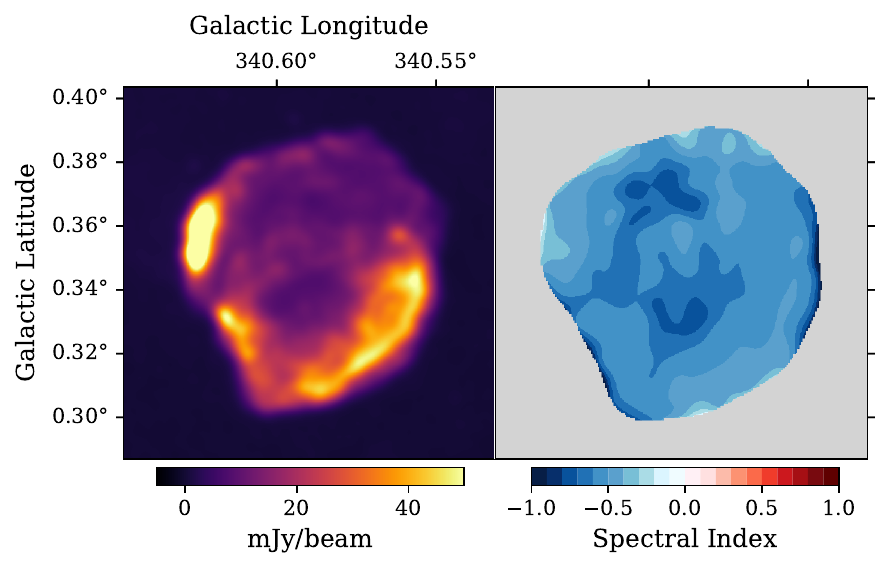}\label{fig:G340.6+0.3}}
    \caption{In-band spectral index maps for known SNRs. The left panels show the 943 MHz Stokes I images from EMU. The right panels show the spectral index maps produced using the four 72 MHz wide sub-bands.}
    \label{fig:snr_maps2}
\end{figure*}

\paragraph{G311.5$-$0.3}
This source is a small bright shell-type SNR. The index map is very smooth, showing little change in index across the source (Figure~\ref{fig:G311.5-0.3}). The three in-band indices are all slightly steeper than the broadband value but still consistent with the expected result for a shell-type SNR. The T-T plot value is most similar to the broadband index, followed by the sub-band value.

\paragraph{G319.9$-$0.7}
This source is a small bow-shock PWN. It does not yet appear in the SNR catalogue \citep{Green_cat}, but it has been studied in depth in the radio and has been confirmed to be polarized \citep{Ng2010,Ball2025}. The index map shows a flatter index along the southwestern edge that steepens toward the northeastern edge (Figure~\ref{fig:G319.9-0.7}). The sub-band index is the most similar to the broadband value, though the T-T plot index is also consistent. The Taylor term method produces a much steeper index for this source.

\paragraph{G326.3$-$1.8}
This SNR is a bright composite source where both the shell and PWN are clearly visible. From the index map, we can see that while the index for the PWN seems to be consistent with the expected value range, there is clearly a problem with the index for the shell (Figure~\ref{fig:G326.3-1.8}). The index for the shell is generally very steep ($\alpha$$<$-1) with some small-scale fluctuations where the index is positive. The steep index of the shell can likely be attributed to missing flux due to missing short spacings as the source is large and the shell emission is diffuse. The total power image also shows that the source sits in a large negative bowl, an imaging artifact that is a consequence of its brightness. These results illustrate how T-T plots may be more reliable for large sources. The sub-band and Taylor term indices are much steeper than both the expected value for an SNR and the broadband value obtained for this source. The T-T plot index is similar to both the expected value range and the broadband index. The T-T plots also show the change in index across the source. In the T-T plot in Figure~\ref{fig:G326.3-1.8_tt}, we show the distinction between the shell and PWN components in the brighter fluxes. It is also notable that the intercept of the trend line is significantly offset from the origin, likely as a consequence of the instrumental effects.

\paragraph{G328.4+0.2}
This SNR is a very bright source that is believed to be a PWN located on the outer edge of the Galaxy \citep{Gaensler2000,Gelfand2007}. The index map shows a flatter index in the central bar region that decreases radially, though not uniformly, across the source (Figure~\ref{fig:G328.4+0.2}). We observe the southern half of the remnant to be steeper than the northern half. The sub-band and Taylor term indices are both much steeper than the broadband value, which is surprising given the small size and brightness of the source, though these values are still consistent with expectations for an SNR. The T-T plot value is the closest to the broadband value. We provide a sample T-T plot in Figure~\ref{fig:G328.4+0.2_tt}. In the T-T plot we can see the same slight change in index observed in the index map, as the data points corresponding to the fainter fluxes deviate slightly from the trend line.

\paragraph{G340.4+0.4}
This source is a shell-type SNR comprised of several bright filaments that surround fainter diffuse emission. There are fluctuations in the index map that seem to be related to the bright filaments (Figure~\ref{fig:G340.4+0.4}). This may be because the flux is fully recovered for the small-scale features but not the more diffuse emission. We observe the index to be flatter along the northern half of the remnant, steepening toward the southern end, which corresponds with the faintness of the emission. All of the in-band indices are steeper than the broadband value but they are also all still consistent with the expected range of values for an SNR. The T-T plot value is the closest to the broadband index.

\paragraph{G340.6+0.3}
This source is a small shell-type SNR. The index map is relatively smooth across the source with some small pockets where the index is slightly steeper (Figure~\ref{fig:G340.6+0.3}). These pockets are once again associated with the fainter parts of the source. There are also several small \ion{H}{2} regions that overlap with this source, which may have a flattening effect on the overall index though this effect is not obviously visible in the index map. The sub-band and Taylor term indices are steeper than the broadband value, but still consistent with the expected index for an SNR. The T-T plot index is consistent with the broadband value. 
\end{document}